\documentclass{emulateapj}

\usepackage{graphicx}
\usepackage{apjfonts}

\slugcomment{ApJ in press}

\begin{document}
\title{Three Dimensional Compressible Hydrodynamic Simulations of Vortices
in Disks}

\shorttitle{Vortices in 3D Hydrodynamic Disks}

\shortauthors{SHEN, STONE \& GARDINER}

\author{Yue Shen, James M. Stone and Thomas A. Gardiner}

\affil{Department of Astrophysical Sciences,
Princeton University, Princeton NJ 08544\\}

%Nov 16, 2005 (Wed) at Peyton Hall -- Y. Shen
%Nov 26, 2005 (Sat) at Peyton Hall -- Y. Shen
%Nov 29, 2005 (Tue) at Peyton Hall -- Y. Shen
%Dec 3, 2005 (Sat) at GC -- Y. Shen
%Dec 6, 2005 (Tue) at Peyton -- Y. Shen
%Feb 16, 2006 (Thu) at Peyton & GC -- Y. Shen (modified version from Jim)

\begin{abstract}

We carry out three-dimensional, high resolution (up to
$1024^2\times 256$) hydrodynamic simulations of the evolution of
vortices in vertically unstratified Keplerian disks using the
shearing sheet approximation.  The transient amplification of
incompressible, linear amplitude leading waves (which has been
proposed as a possible route to nonlinear hydrodynamical
turbulence in disks) is used as one test of our algorithms; our
methods accurately capture the predicted amplification, converges
at second-order, and is free from aliasing. Waves expected to
reach nonlinear amplitude at peak amplification become unstable to
Kelvin-Helmholtz modes when
%$k_x \delta v_y \gtrsim \Omega$ (where
%$k_x$ is the radial wavenumber of the perturbation, $\delta v_y$
%the azimuthal component amplitude, and $\Omega$ the angular
%velocity).
$\mid W_{\rm max}\mid\gtrsim \Omega$ (where $W_{\rm max}$ is the
local maximum of vorticity and $\Omega$ the angular velocity). We
study the evolution of a power-law distribution of vorticity
consistent with Kolmogorov turbulence; in two-dimensions
long-lived vortices emerge and decay slowly, similar to previous
studies. In three-dimensions, however, vortices are unstable to
bending modes, leading to rapid decay. Only vortices with a length
to width ratio smaller than one survive; in three-dimensions the
residual kinetic energy and shear stress is at least one order of
magnitude smaller than in two-dimensions.  No evidence for
sustained hydrodynamical turbulence and transport is observed in
three-dimensions. Instead, at late times the residual transport is
determined by the amplitude of slowly decaying, large-scale
vortices (with horizontal extent comparable to the scale height of
the disk), with additional contributions from nearly
incompressible inertial waves possible. Evaluating the role that
large-scale vortices play in astrophysical accretion disks will
require understanding the mechanisms that generate and destroy
them.

\end{abstract}

\keywords{accretion, accretion disks -- solar system: formation}

\section{INTRODUCTION}

Many aspects of the structure and evolution of astrophysical
accretion disks are controlled by the mechanism for angular
momentum transport. When the disk is dynamically coupled to a weak
magnetic field, the magnetorotational instability (MRI, e.g.,
Balbus \& Hawley 1998) generates MHD turbulence which is a
vigorous source of angular momentum transport (for a recent
review, see Balbus 2003). However, in neutral or very weakly
ionized disks (e.g., proto-planetary disks, and dwarf nova systems
in quiescence) the magnetic field may be decoupled from the gas,
and the MRI may not operate (Salmeron \& Wardle 2005; Fromang, Terquem,
\& Balbus 2002; Gammie \& Menou 1998).  It these cases, it is important to
investigate purely hydrodynamical mechanisms for the transport of
angular momentum.

Although Keplerian shear flows are linearly stable according to
the Rayleigh criterion, it has long been supposed that there
exists some mechanism that produces turbulence and transport via
nonlinear instabilities at high Reynolds number. To date, however,
no mechanism for such instabilities has been explicitly
demonstrated, and no evidence for self-sustained hydrodynamical
turbulence has been found in direct numerical simulations of
Keplerian disks at the same Reynolds number needed to capture
known nonlinear instabilities in Cartesian shear flows (Balbus,
Hawley, \& Stone 1996; Hawley, Balbus \& Winters 1999).  Even if
nonlinear instabilities exist in Keplerian shear flows, Lesur \&
Longaretti (2005) argue that current simulations already show the
associated transport would be negligable.  Recently, new interest
has focused on the role that transient growth of incompressible,
leading waves might play in generating hydrodynamical turbulence
(Chagelishvili et al. 2003; Umurhan \& Regev 2004; Yecko 2004).
The amplification factor of incompressible (vortical) waves is of
order $(k_{x,0}/k_y)^2$, where $k_{x,0}$ and $k_y$ are the initial
wavenumbers in the radial and azimuthal directions respectively
(Johnson \& Gammie 2005a, hereafter JGa). For leading waves that
are initially tightly wound $\mid k_{x,0}\mid/k_y \gg 1$, large
amplification factors are possible, potentially leading to
nonlinear amplitudes. This has been suggested as one step in the
route to self-sustained hydrodynamical turbulence in disks
(Afshordi et al 2005; Mukhopadhyay et al. 2005; but see Balbus \&
Hawley 2006). One purpose of this paper is to investigate the
effect of transient growth of leading waves in fully
three-dimensional (3D) compressible hydrodynamical disks.

On the other hand, there is continued interest in the role that
vortices play in disks.  There is particular interest in the context of
proto-planetary disks, where it has been suggested that vortices play a
role in trapping dust and aiding in grain-growth (e.g., Barge \& Sommeria
1995; Adams \& Watkins 1995; Johansen, Anderson, \& Brandenburg 2004).
In addition, if vortices produce outward angular momentum transport
they may play an important role in the structure and evolution of
proto-planetary disks as well as other weakly ionized accretion disks
(e.g., Li et al. 2001; Barranco \& Marcus 2005, hereafter BM; Johnson
\& Gammie 2005b, hereafter JGb).  A variety of authors have studied
the evolution of random vorticity in both 2D and 3D, and in both
incompressible and compressible gases.  Long-lived, anticyclonic (with
rotation opposite to the background shear) vortices have been found in
2D Keplerian disks (e.g., Godon \& Livio 1999, 2000; Umurhan \& Regev
2004; JGb).  The work of JGb is particularly relevant to this study in
that they considered the evolution of a spectrum of vorticity consistent
with Kolmogorov turbulence in fully compressible 2D simulations.
In every case, they found the kinetic energy and transport decays from
the initial conditions (sustained turbulence was not evident), although
the rate of decay was affected by factors such as the horizontal extent
of the domain.  These authors argued that compressibility significantly
increases the outward angular momentum transport during the decay.
Fully 3D simulations of the evolution of vorticity in stratified disks in
the anelastic approximation have been presented by BM; they found vortices
survive (and in fact, are produced) in the upper regions of the disk above
a few scale heights.  In the midplane, three-dimensional instabilities
destroy vortices and reduce transport.  Here we extend the work of JGb
and BM by studying vortices in fully {\em compressible} unstratified
disks in 3D. We will report on 3D studies of stratified disks elsewhere.

This paper is organized as follows. In Section 2 we describe the numerical
methods and shearing sheet model used in our simulations. As a test,
we study the evolution of both incompressible and compressible planar
shearing waves in two-dimensions in section 3; we point out
that at large amplitude vortical waves become Kelvin-Helmholtz
unstable.  In section 4 we discuss the evolution of random
vortical perturbations in both 2D and 3D. We summarize and conclude in
section 5.

\section{METHOD}

To study the local hydrodynamics of Keplerian accretion disks, the
{\it shearing sheet} approximation is adopted (Hawley, Gammie, \& Balbus 1995),
that is the
equations of motion are solved in a Cartesian frame centered at a
radius $R_0$ in the disk, with linear extent much less than $R_0$,
and which corotates at angular velocity $\Omega(R_0)$ .
In this frame, the Cartesian $x$
coordinate corresponds to radius and $y$ to azimuth. The
hydrodynamic equations are expanded to lowest order in $H/R_0$ in
this frame ($H\equiv c_s/\Omega$ is the local scale height of the
disk, $c_s$ is the isothermal sound speed), giving
\begin{equation}\label{continuity}
\frac{\partial \rho}{\partial t}+\nabla\cdot(\rho \mathbf{v})=0\ ,
\end{equation}
\begin{equation}\label{Euler}
\frac{\partial {\mathbf v}}{\partial t}+{\mathbf v}\cdot\nabla
{\mathbf v}=-\frac{1}{\rho}\nabla p-2{\mathbf \Omega}\times
{\mathbf v}+2 q\Omega^2x\hat{{\mathbf x}}\ ,
\end{equation}
\begin{equation}\label{eos}
p=c_s^2\rho\ ,
\end{equation}
where $q\equiv -d\ln\Omega/d\ln R$ is the shear parameter.  For
Keplerian disks $q=3/2$.  An isothermal equation of state is used
throughout this paper.  In Euler's equation (\ref{Euler}), the
term $-2{\bf \Omega}\times {\bf v}$ is the Coriolis force in the
rotating frame, and $2q\Omega^2x\hat{{\bf x}}$ represents the
effect of tidal gravity.  The latter is added as the gradient of an
effective potential, which combined with the fact we use a numerical
method which solves the equations in conservative form (see below),
guarantees the net radial and azimuthal momentum is conserved.
Vertical gravity is neglected for the
unstratified disks studied here.

We have omitted an explicit viscosity in our ideal hydrodynamics
formalism.  We show in the next section that the numerical dissipation
inherent in our scheme acts to damp short wavelength
perturbations as desired (in particular, it prevents aliasing of
trailing into leading waves), while having little effect on
well-resolved modes (more than 16 grid points per wavelength).  It
is difficult to define an effective Reynolds number for our calculations,
since the numerical dissipation is a steep function of resolution.
We show below the amplitude errors in the evolution of shearing
waves converges at better than second-order with our methods.

Our calculations begin from an equilibrium state consisting of
constant density $\rho=\rho_0$, pressure $p=p_0=c_s^2 \rho_0$, and
purely azimuthal velocities corresponding to Keplerian rotation
${\bf v}={\bf v}_0=-q\Omega x\hat{y}$.  We choose
$c_s=\Omega=10^{-3}$ (which implies $H=1$), and $\rho_0=1$. Our
computational domain is of size $L_x \times L_y \times L_z$, with
$L_x = L_y =4H$ and $L_z=H$.  In 2D runs, our domain represents
the equatorial plane and is of size $L_x \times L_y=4H\times 4H$
(except in some 2D tests where
$L_x=L_y=0.5H$, see \S\S 3.2 and 3.3). Note the difference
in the azimuthal velocity across the horizontal domain ($6c_s$)
exceeds the sound speed. JGb found that in the evolution of random
vortical perturbations in 2D, the amplitude and rate of decay of
the resulting transport depends on the size of this velocity
difference (and therefore the size of the domain). We adopt the
fiducial model studied by JGb and expect, as in 2D, that the
effect of compressibility will be reduced in smaller domains than
studied here.

Our simulations use the 3D version of the Athena code (Gardiner \&
Stone 2005; 2006). Athena is a single-step, directionally unsplit
Godunov scheme for hydrodynamics and MHD which uses the piecewise
parabolic method (PPM) for spatial reconstruction, a variety of
approximate Riemann solvers to compute the fluxes, and the corner
transport upwind (CTU) unsplit integration algorithm of Colella
(1990). The calculations presented here use fluxes computed using
the two-shock approximation
for isothermal hydrodynamics.  The algorithms are second
order accurate for all hydrodynamic and MHD wave families
(Gardiner \& Stone 2006); we demonstrate better than second-order
convergence for plane waves in the shearing sheet in \S3, indicating
that our implementation of the shearing sheet boundary conditions
in Athena is at least second-order accurate.

The computational domain is divided into a grid of $N_x \times N_y
\times N_z$ zones.  We choose $N_x = N_y \equiv N$ and $N_z =
N/4$. We study the convergence of our numerical results using
grids of size $N=32$ to $N=2048$ in two-dimensions, and $N=128$ to
$N=1024$ in three-dimensions.

Important diagnostics of the flows considered here are the
dimensionless angular momentum flux in fluctuations, defined as
\begin{equation}
\alpha\equiv \langle\rho v_x v_y^{\prime}\rangle /(\rho_0c_s^2)\ ,
\end{equation}
where $\langle\ \rangle$ denotes a volume average, and
$v_y^{\prime} = v_y+q\Omega x$ is the difference in the azimuthal
velocity compared to Keplerian rotation.  Tests of our numerical methods
using epicyclic waves show the mean radial flux $\langle \rho v_x \rangle$
remains zero over hundreds of shearing times, thus there is no need to
correct equation (4) for any mean radial motion.
The volume averaged perturbed kinetic energy density,
including only motions in the horizontal plane, is
\begin{equation}
E_{{\rm K}}=\langle\frac{1}{2}\rho(v^{2}_x+v_y^{\prime 2})\rangle\
.
\end{equation}
Finally, the $z-$component of vorticity
\begin{equation}
W_z = \frac{\partial v_y}{\partial x} - \frac{\partial
v_x}{\partial y}
\end{equation}
is an important diagnostic.  In 2D incompressible flows, it is exactly
conserved (i.e., the Lagrangian derivative $dW_z/dt$ is zero). In 2D,
smooth, compressible flows, the potential vorticity $\tilde{W_z}\equiv
(W_z+2\Omega)/\rho$ is conserved (JGa; the extra $2\Omega$ term comes
from the Coriolis force in the shearing sheet formalism) if the flow is
adiabatic and isentropic.  However, in 2D flows with shocks both vorticity
and potential vorticity can be generated through shock-curvature and
shock-shock interactions (Kevlahan 1997).  In 3D, $W_z$ can evolve
freely. We track both $W_z$ and $\tilde{W_z}$ as a diagnostic of our
methods, and of the effects of shocks.

\begin{figure*} \centering
\includegraphics[width=0.45\textwidth]{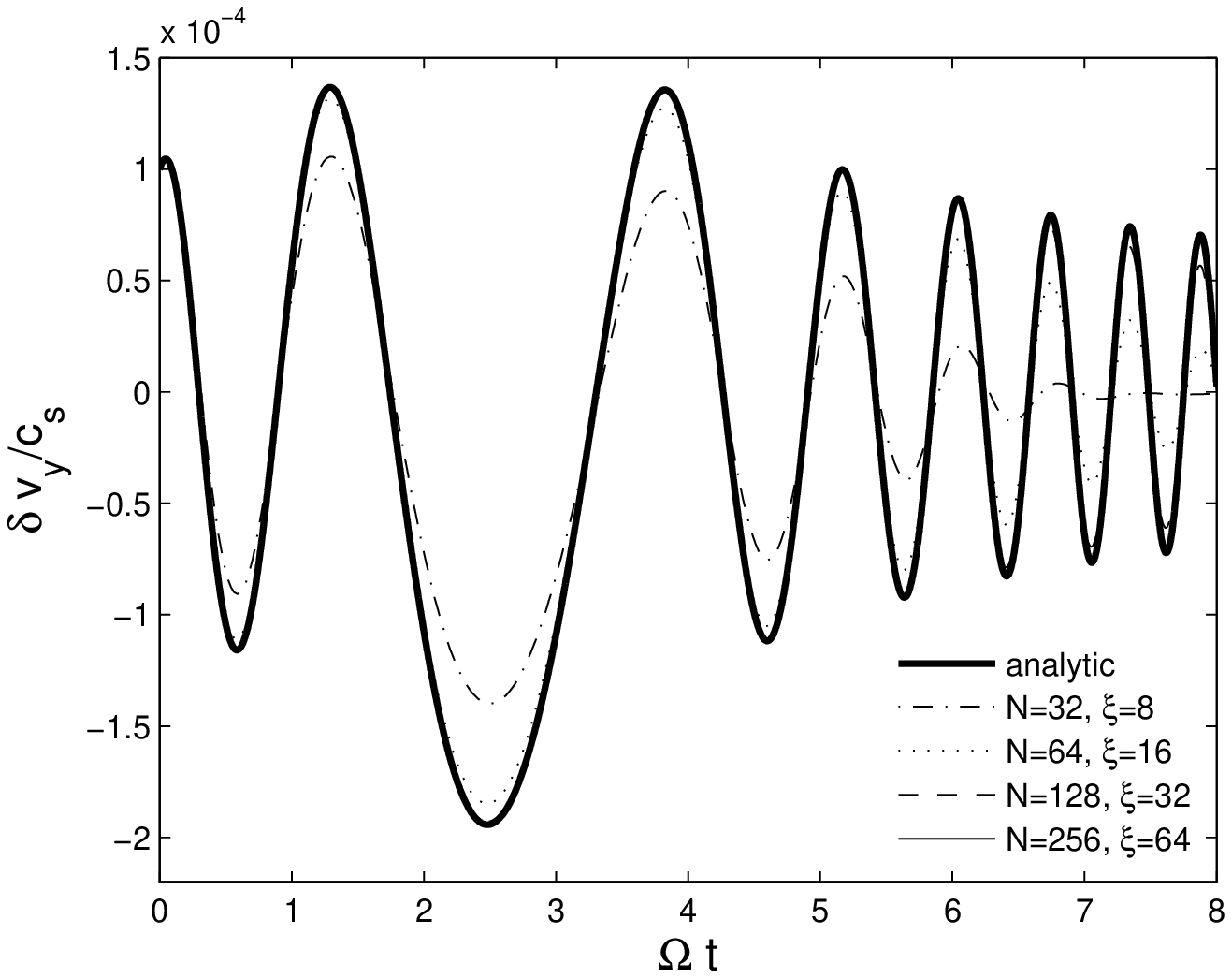}
\includegraphics[width=0.45\textwidth]{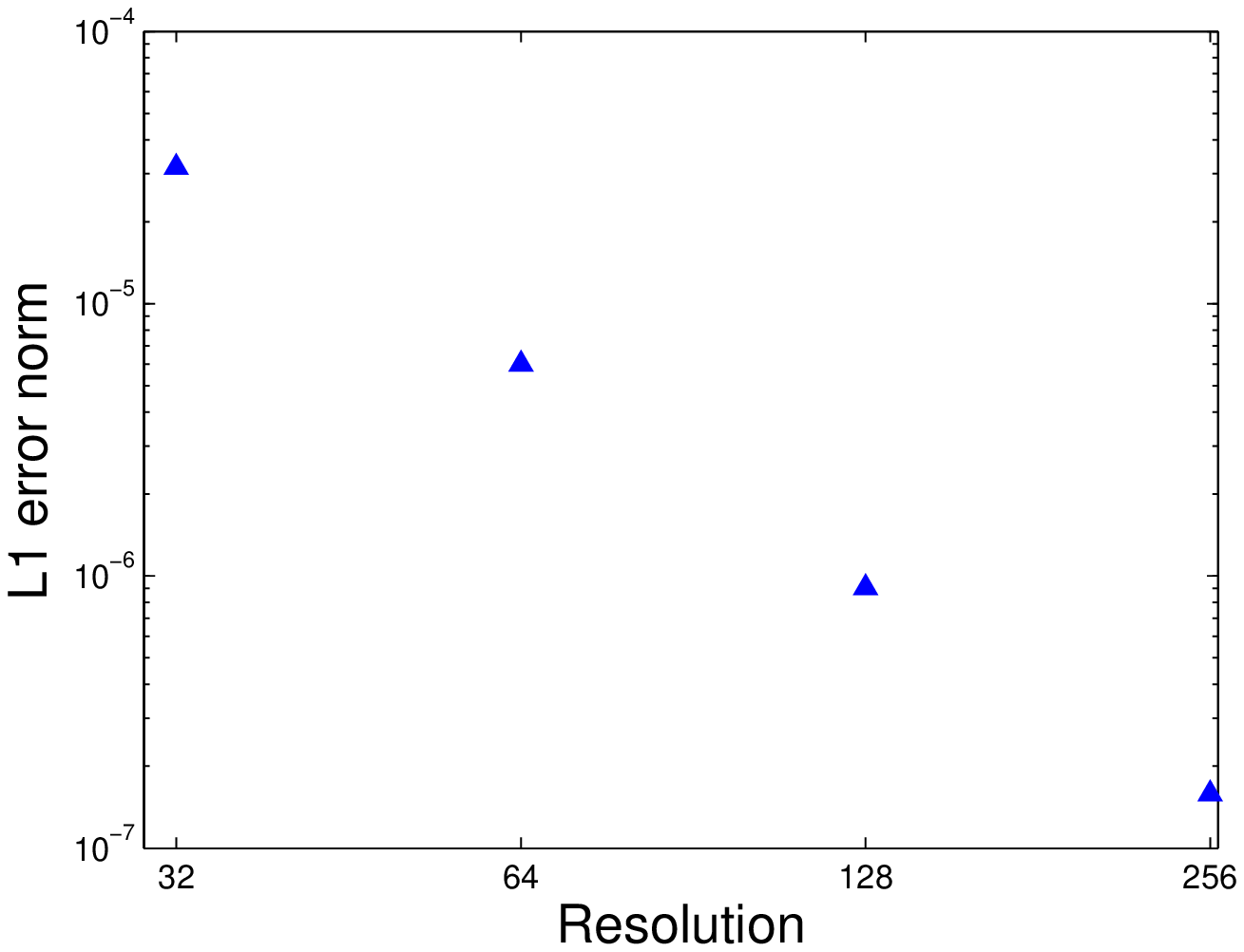}
\caption{Evolution of a linear amplitude, compressible, shearing
wave. {\em left:} The analytic solution (heavy solid line) for the
wave amplitude is compared to numerical simulations at various
grid resolutions $N \times N$, where $\xi$ is the number of grid
points per wavelength in the $x-$direction in the initial
conditions. {\em right:} L1 error norm in the $y$-component of the
velocity summed over $0 \leq \Omega t \leq 6$ versus resolution.
The error is converging at $N^{-2.5}$. } \label{figNV}
\end{figure*}

\section{TWO-DIMENSIONAL EVOLUTION OF SHEARING WAVES}

We begin by using the linear theory of plane waves
in the shearing sheet as a test of our numerical
algorithms.  Our numerical methods then allow us to investigate
what happens when leading waves are amplified to nonlinear
amplitudes.  We present the evolution of random vorticity fields
in \S4.

In the two-dimensional
($x-y$) shearing sheet, velocity perturbations
representing plane waves have the form
\begin{equation}
%\begin{split}
v_{x} = \delta v_{x}(t)\exp[ik_x(t)x+ik_yy]\ ,\\
v_{y}+q\Omega x = \delta v_{y}(t)\exp[ik_x(t)x+ik_yy]
%\end{split}
\end{equation}
where
\begin{equation}
 k_x(t)=k_{x,0}+q\Omega k_yt
\end{equation}
and $k_{x,0}$ and $k_y>0$ are constant.  The subscript 0 denotes
quantities at the initial time $t=0$.  The background shear causes
the $x-$component of the wavevector to evolve linearly with time.
%The evolution of linear amplitude shearing waves can be treated
%analytically in the short-wavelength %or WKB
%limit (i.e., $Hk_y\gg 1$).
Solutions can be divided into both compressible ($\nabla \cdot
{\bf v}\neq 0$) and incompressible ($\nabla \cdot {\bf v}=0$)
forms and treated analytically in the short-wavelength (i.e.,
$Hk_y\gg 1$), low-frequency ($\partial_t \ll c_sk_y$) limit (e.g.
JGa). We consider each separately below.

\subsection{Compressible plane waves}

%In the WKB limit,
The amplitude of compressible velocity perturbations evolve
according to (JGa)
\begin{equation}\label{dVy}
\ddot{\delta v}_{y}+(c_s^2k^2+\kappa^2)\delta v_y=0\ ,
\end{equation}
where $\kappa^2\equiv 2(2-q)\Omega^2$ is the square of epicyclic
frequency.  The solution of (\ref{dVy}) can be written in terms of
parabolic cylinder functions, see JGa for details.  Note that
equation (\ref{dVy}) is valid for all wavelengths and frequencies
as long as the perturbed potential vorticity is zero (JGa).

Figure \ref{figNV} ({\em left}) plots the evolution of a
compressible shearing wave with initial wavenumbers
$k_{x,0}=-4(2\pi/L_x)$ and $k_{y}=2\pi/L_y$ with $L_x=L_y=4H$, and
initial amplitude $\delta v_{x,0}/c_s=4\times10^{-4}$.  Also shown
is the amplitude of the wave measured from numerical simulations
at resolutions of $N=32,\ 64,\ 128,$ and 256 grid points per
horizontal dimension.  An important measure of the numerical
resolution for plane waves is the number of grid points per
initial wavelength $\lambda_{x,0} = 2\pi/k_{x,0}$ in the
$x-$direction,
\begin{equation}
\xi \equiv \frac{\mid\lambda_{x,0}\mid}{\Delta x}=\frac{2\pi
N}{\mid k_{x,0} \mid L_x}
\end{equation}
Thus, $N=32,\ 64,\ 128,$ and 256 corresponds to $\xi=8,\
16,\ 32,$ and 64 grid points per initial wavelength respectively.

The oscillatory behavior of the amplitude of a compressible
shearing wave is accurately traced by the numerical results. There
is no phase error at any resolution, and at $\xi=16$ grid points
per wavelength the peak amplitude of the oscillations is 0.96 of
the analytic value in the first peak, and this value increases
rapidly with resolution. Below this resolution the waves are
smoothly damped.  As a quantitative measure of the errors, figure
\ref{figNV} ({\em right}) plots the L1 error norm in the
$y$-component of the velocity summed over the interval $0 \le
\Omega t \le 6$ as a function of numerical resolution. The slope
of the line is -2.5, indicating our numerical solution is
converging at better than second-order. This is direct evidence
that our implementation of the shearing sheet boundary conditions,
which are required to capture the evolution of shearing waves, is
no less accurate than our integration algorithm.

\subsection{Incompressible plane waves}

\begin{figure*} \centering
\includegraphics[width=0.45\textwidth]{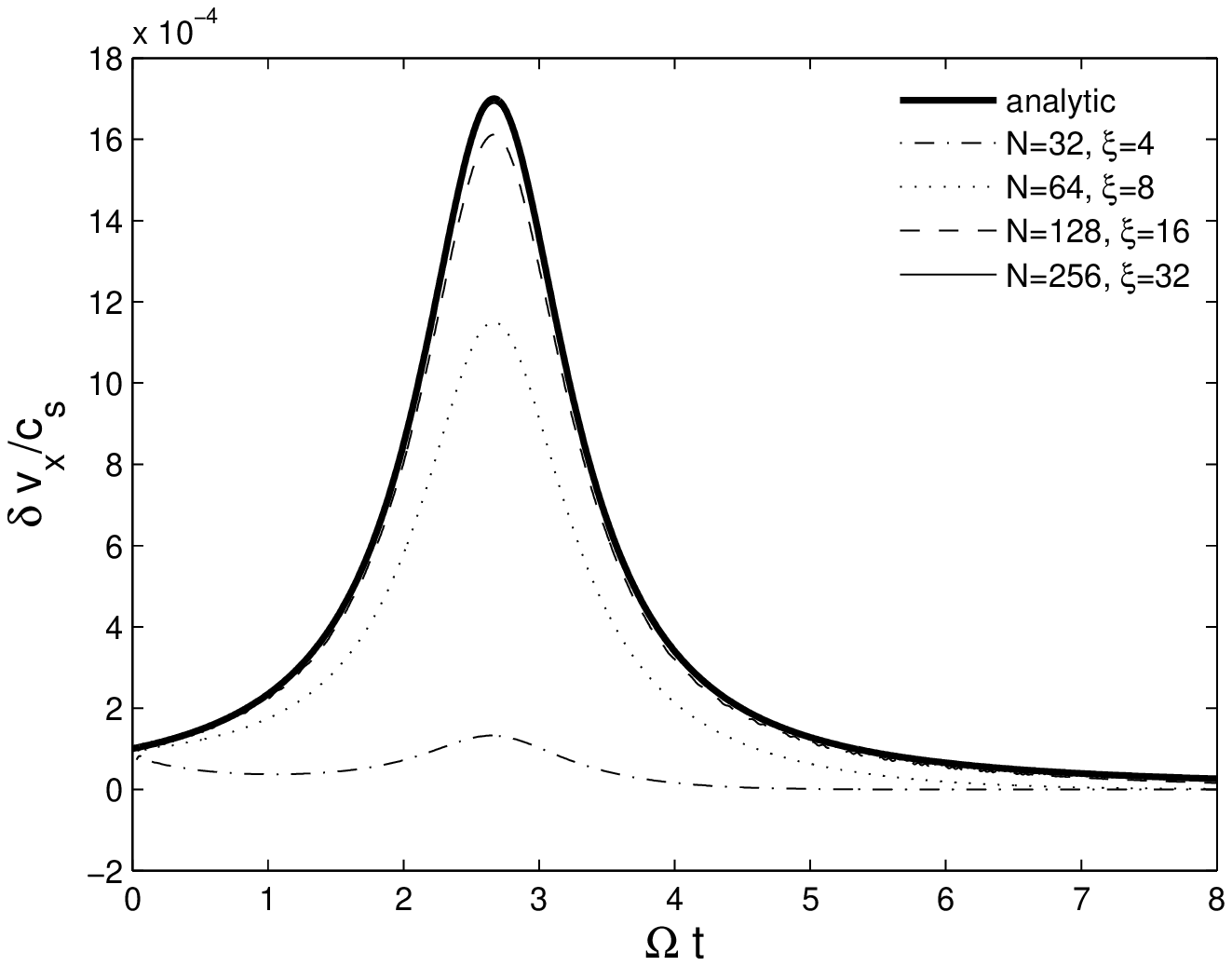}
\includegraphics[width=0.45\textwidth]{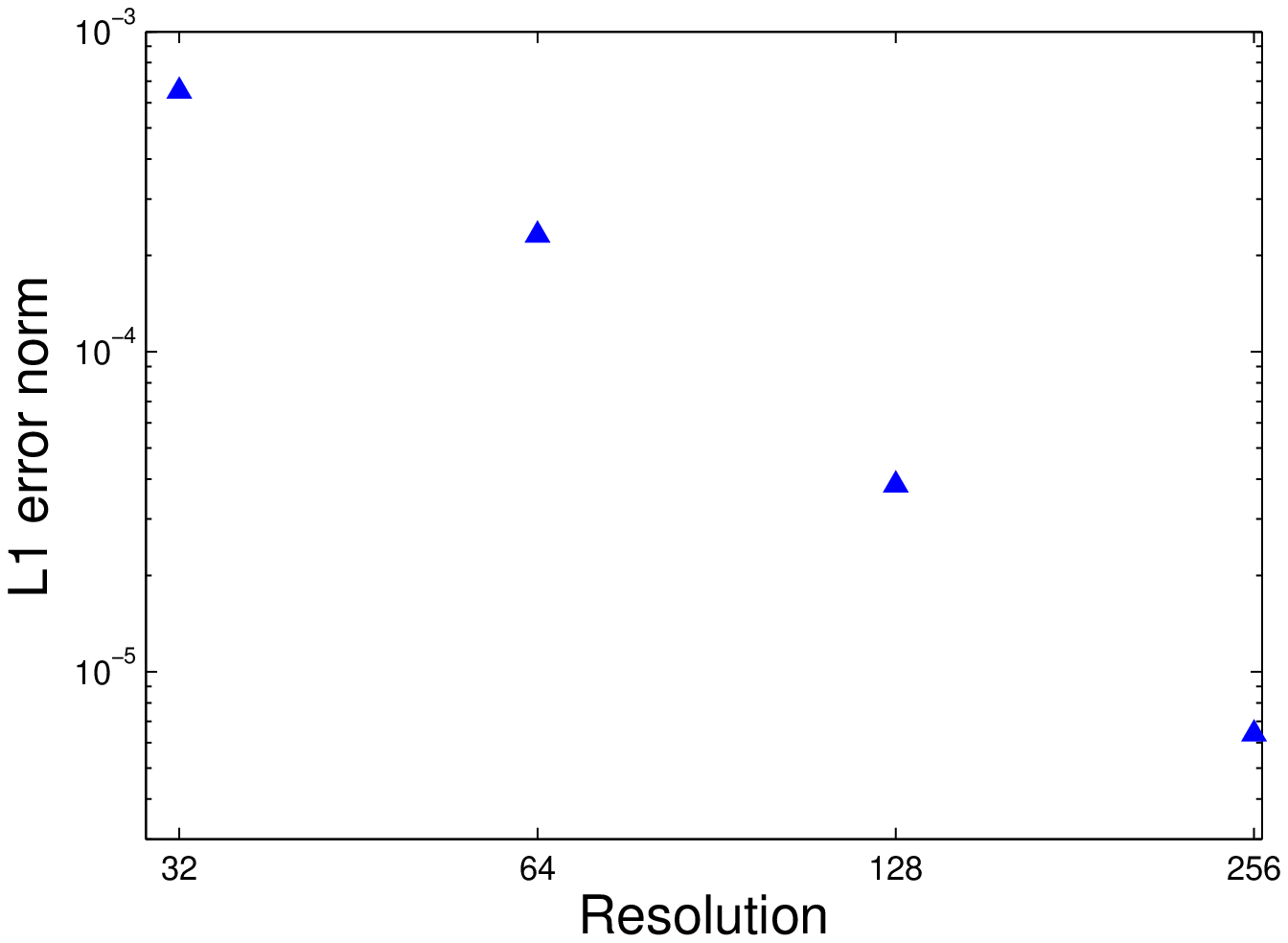}
\caption{Evolution of a linear amplitude, incompressible, shearing
wave. {\em left:} The analytic solution (heavy solid line) for the
wave amplitude is compared to numerical simulations at various
grid resolutions $N \times N$, where $\xi$ is the number of grid
points per wavelength in the $x-$direction in the initial
conditions. {\em right:} L1 error norm in the $x$-component of the
velocity summed over $0 \leq \Omega t \leq 6$ versus resolution.
The error is converging at $N^{-2.5}$ for $N\ge 64$. }
\label{figVor}
\end{figure*}

For vortical (incompressible) shearing waves in the
short-wavelength, low-frequency limit, the amplitudes of the
velocity perturbation evolve as
\begin{equation}\label{dVx}
\delta v_x=\delta v_{x,0}\frac{k_0^2}{k^2}\ ,\qquad \delta
v_y=-\frac{k_x}{k_y}\delta v_x\ ,
\end{equation}
where $k^2=k_x^2+k_y^2$.  If the vortical wave starts as leading,
i.e., $k_{x,0}<0$, it will swing to trailing at a time $t_{\rm
max}= \mid k_{x,0}\mid/(q\Omega k_y)$. During this time, the
amplitude of the $x-$ component velocity perturbation is
temporarily amplified, attaining a maximum value $\delta v_{x,{\rm
max}}={\cal A}\delta v_{x,0}$ at $t=t_{\rm max}$, where
\begin{equation}
{\cal A}\equiv 1+(k_{x,0}/k_y)^2
\end{equation}
is the peak amplification factor (e.g., Chagelishvili et al. 2003;
Umurhan \& Regev 2004; JGa). Note that ${\cal A}$ is also the peak
amplification factor of the kinetic energy in the wave.
% The amplitude of the $y-$ component velocity perturbation doesn't evolve
% as \delta v_x; instead, \delta v_y=0 at $t_{\rm max}$. But the combination
% of \delta v_x^2 + \delta v_y^2 is still amplified during the leading to
% trailing transition. That's the reason why the amplification factor of
% kinetic energy is also A, not A^2.

Figure \ref{figVor} ({\em left}) plots the evolution of an
incompressible shearing wave with initial wavenumbers
$k_{x,0}=-8(2\pi/L_x)$ and $k_{y}=2(2\pi/L_y)$, and initial
amplitude $\delta v_{x,0}/c_s=10^{-4}$. A smaller computational
domain is used in this test, with $L_x=L_y=0.5H$, to minimize the
effects of compressibility.  To ensure the numerical
representation of the perturbation satisfied the centered
difference form of $\nabla \cdot {\bf v} = 0$ initially, the plane
wave solution is used to compute the analytic form for the
$z-$component of a velocity potential $A_z$ which satisfies ${\bf
v} = \nabla \times {\bf A}$, and then centered differences of $v_x
=
\partial A_z / \partial y$ and $v_y = -
\partial A_z / \partial x$ are used to compute the initial velocities on the
grid.  Also shown is the evolution of the wave amplitude measured
from numerical simulations at resolutions of $N=32,\ 64,\ 128,$
and 256 grid points per horizontal dimension, corresponding to
$\xi=8,\ 16,\ 32,$ and 64 grid points per initial wavelength
respectively.

The expected amplification factor for the wave is ${\cal A}=17$.
As in the compressible wave case, we find our numerical solutions
accurately track the analytic solution above $\xi=16$ grid points
per wavelength. Below this value the wave is smoothly damped.
Figure \ref{figVor} ({\em right}) plots the L1 error norm in the
$x$-component of the velocity summed over the interval $0 \leq
\Omega t \leq 6$ versus resolution; once again we find uniform
convergence in the errors at a rate of 2.5 for $N\ge64$, that is
better than second-order.

\begin{figure} \centering
{\includegraphics[width=0.5\textwidth]{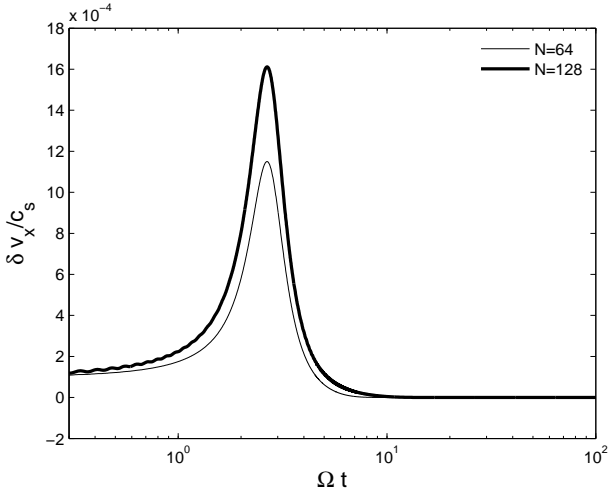}}
\caption{Amplification of the amplitude of an incompressible plane
wave at with $k_{x,0}/k_y=-4$ and resolutions of $64^2$ and
$128^2$. Aliasing, if present, should occur at $q\Omega
t=N_x/n_y-k_{x,0}/k_y$; note it is completely absent at both
resolutions. } \label{FigAliasing}
\end{figure}

An important property of our numerical solutions is the lack of
any aliasing.  As discussed in Umurhan \& Regev (2004) and JGb,
aliasing is dangerous in the sense that it causes power to be
artificially injected by trailing waves into leading waves, which
can subsequently by amplified by the shear to repeat the loop. The
symptom of aliasing is the reoccurrence of transient growth at
times $\Omega t = (N_x/n_y-k_{x,0}/k_y)/q$ (where $n_y=2$ is the
y-direction wave number) during the decay phase of trailing waves
(e.g., fig. 7 in JGb). Figure \ref{FigAliasing} shows the
evolution of two runs with parameters
$(k_{x,0},k_y)=(-16\pi/L_x,4\pi/L_y)$, $\delta
v_{x,0}/c_s=10^{-4}$ and N=64, 128 to $\Omega t=100$. Note that
after the original peak in the amplification, the amplitude decays
monotonically, without any further transient growth.  We postulate
that the lack of aliasing in our method is related to its
dissipation properties. At low resolution (below 16 grid points
per wavelength), Figure \ref{figVor}a shows our method smoothly
damps waves.  Thus, as trailing waves are sheared, our method
damps them before their wavelength drops to the grid resolution
and they are aliased into new leading waves.  Note this does not
mean our methods are overly diffusive, because the dissipation in
Godunov schemes is a strong nonlinear function of resolution.
Thus, for resolved waves $\xi \ge 16$ the dissipation is smaller
than ZEUS-like methods.  In some sense, the dissipation properties
of Godunov methods are ideal for preventing aliasing.

\begin{figure*} \centering
\includegraphics[width=0.45\textwidth]{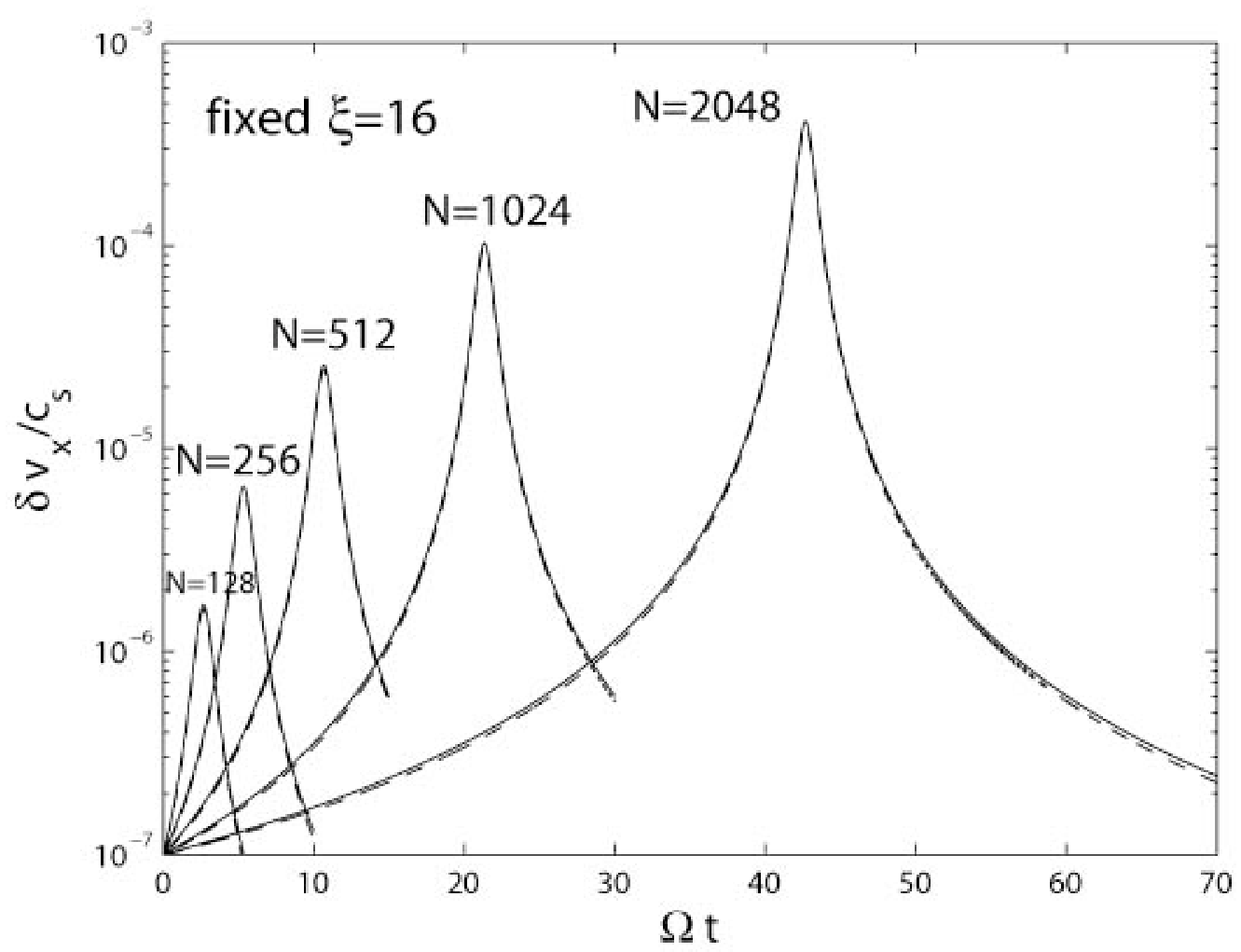}
\includegraphics[width=0.45\textwidth]{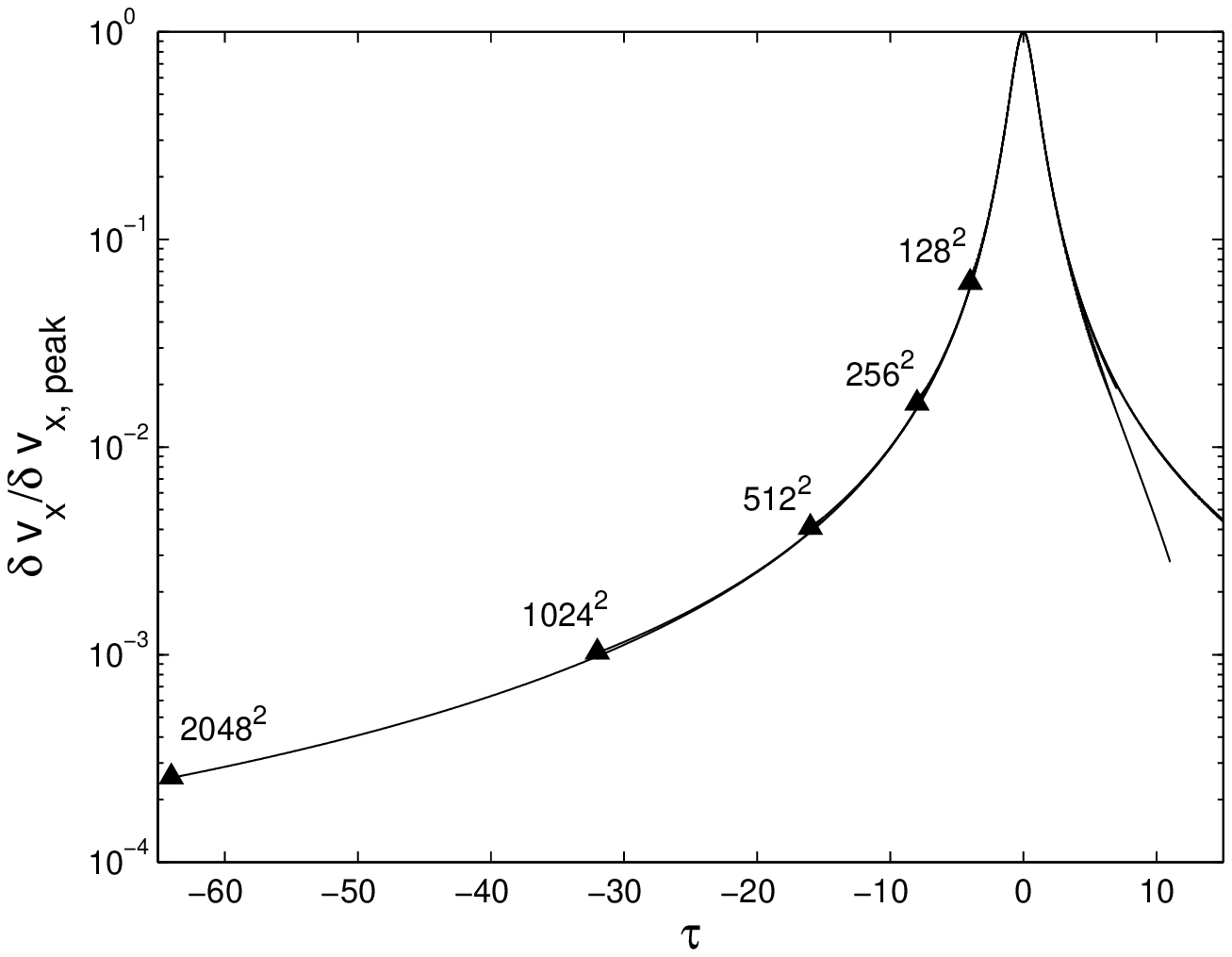}
\caption{Evolution of the amplitude of incompressible (vortical)
plane waves with different initial $\mid k_{x,0}/k_y\mid$ and
fixed numerical resolution $\xi=16$ grid points per wavelength.
{\em left:}  Curves labelled by $N=128, 256, 512, 1024$ and $2048$
have $\mid k_{x,0}/k_y\mid=4, 8, 16, 32,$ and 64 respectively. The
peak amplification is ${\cal A}=17,\ 65,\ 257,\ 1025,\ 4097$. Both
numerical (dashed line) and analytic (solid line) solutions are
shown. {\em right:} Same as {\em left}, but plotted in terms of
$\tau\equiv q\Omega t+k_{x,0}/k_y$. } \label{figAmplification}
\end{figure*}

Using fixed resolution per wavelength $\xi=16$, we have varied the
initial ratio $k_{x,0}/k_y$ ($k_y=4\pi/L_y$ is fixed) to obtain
different amplification factors ${\cal A}$.  To keep the waves
linear at peak amplification, we choose a very small initial
amplitude $\delta v_{x,0}/c_s=10^{-7}$. Using grids of
$N_x=N_y=128,\ 256,\ 512,\ 1024,\ 2048$, we have evolved waves
with $\mid k_{x,0}/k_y \mid =4,\ 8,\ 16,\ 32,$ and 64
respectively, giving peak amplification factors of ${\cal A}=17,\
65,\ 257,\ 1025$, and 4097.  The results are plotted in figure
\ref{figAmplification} ({\em left}) (dashed lines) along with
analytical solutions (solid lines).  In each case, the numerical
solution tracks the analytic accurately. As long as the initial
wave is resolved, our method captures large amplification factors
as well as it does small.  Figure \ref{figAmplification} ({\em
right}) plots the same data, but scales the amplitude to the peak
value, and uses $\tau\equiv q\Omega t+k_{x,0}/k_y$ as the time
coordinate. The start time of each calculation is donated by a
filled triangle. In this case, every curve lies atop one another,
as expected.

\subsection{Kelvin-Helmholtz Instability of Incompressible waves}

\begin{figure*}
  \centering
    \includegraphics[width=0.45\textwidth]{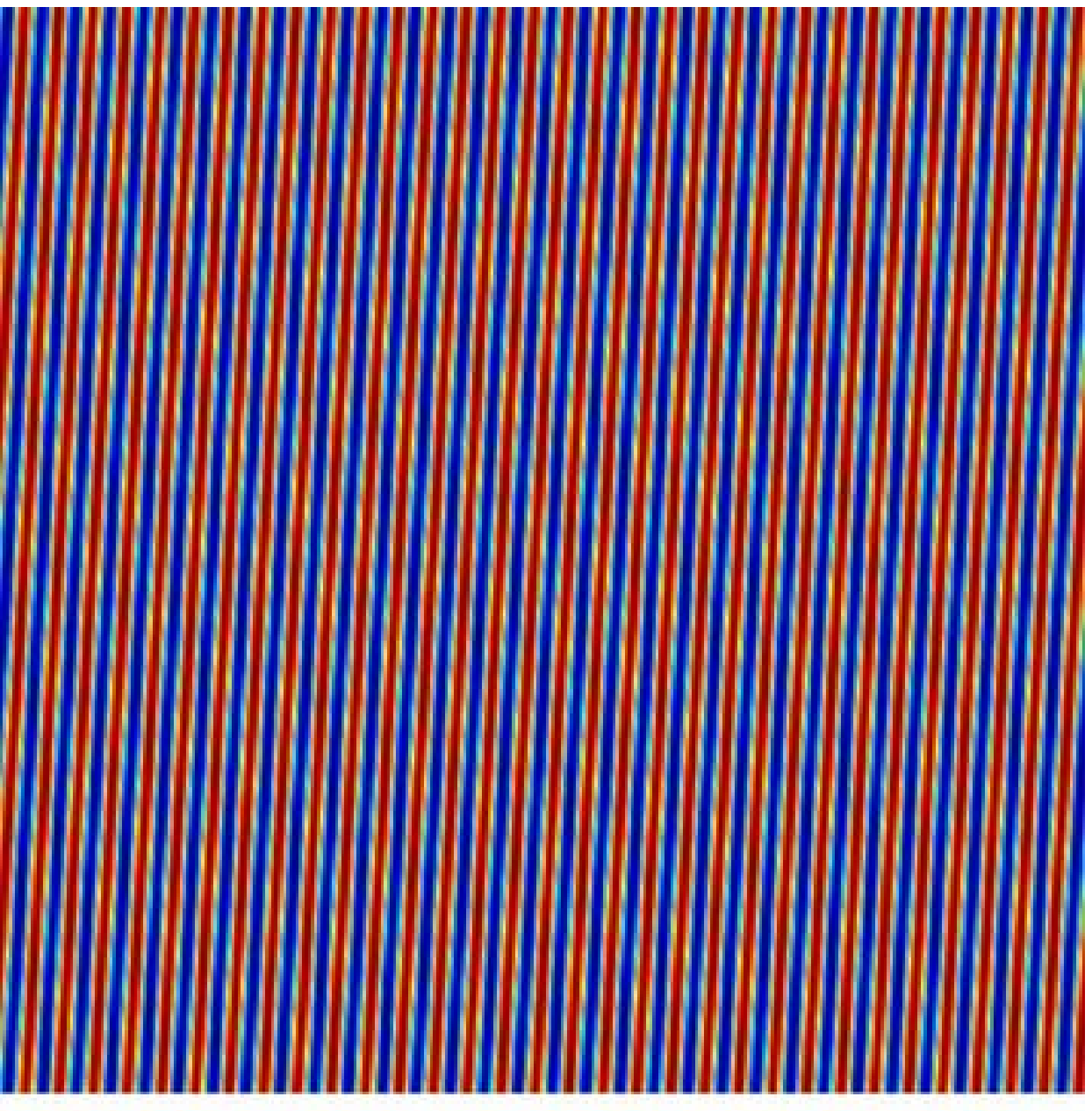}
    \hspace{0mm}
    \includegraphics[width=0.45\textwidth]{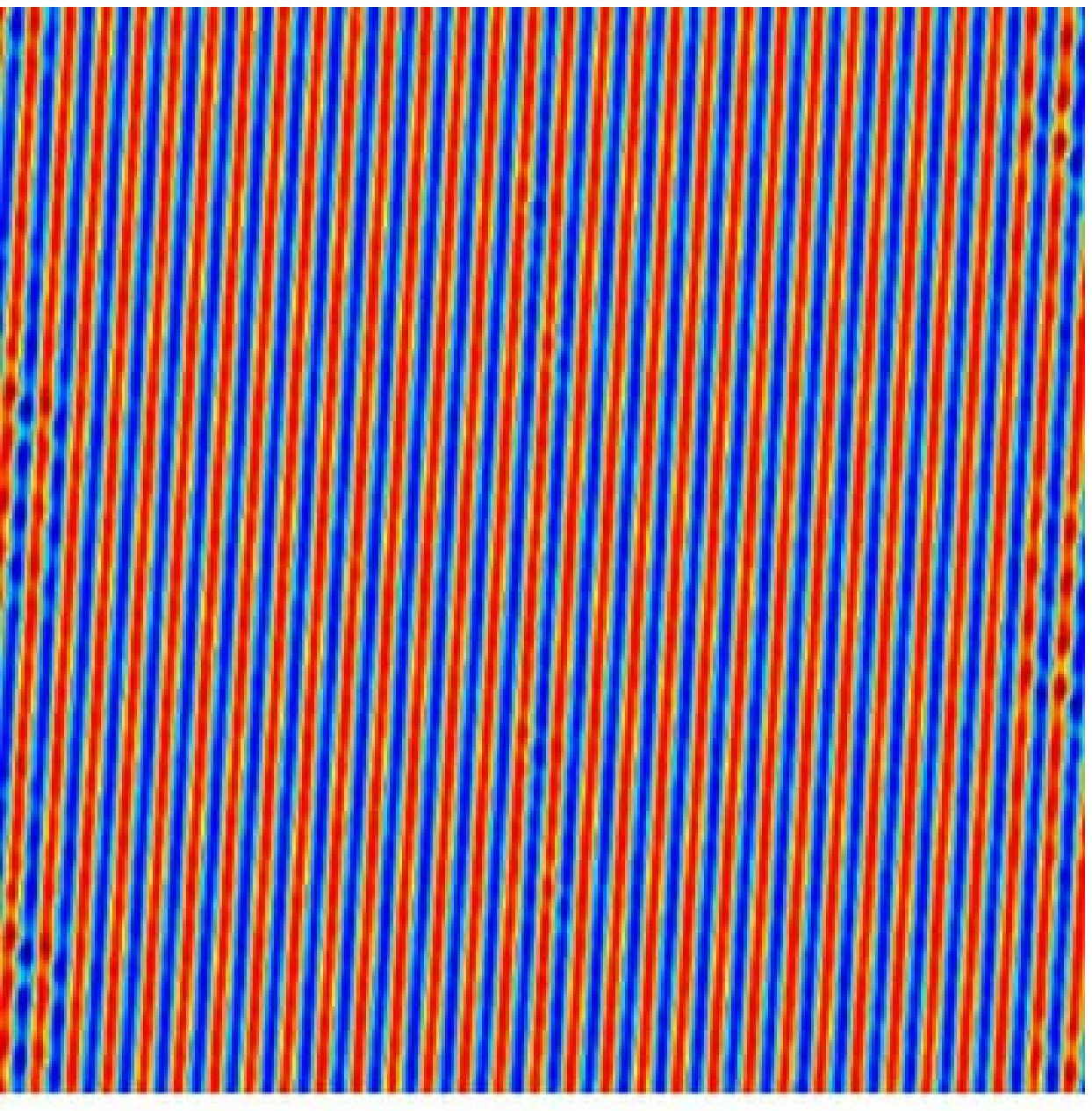}
    \\[0mm]
    \includegraphics[width=0.45\textwidth]{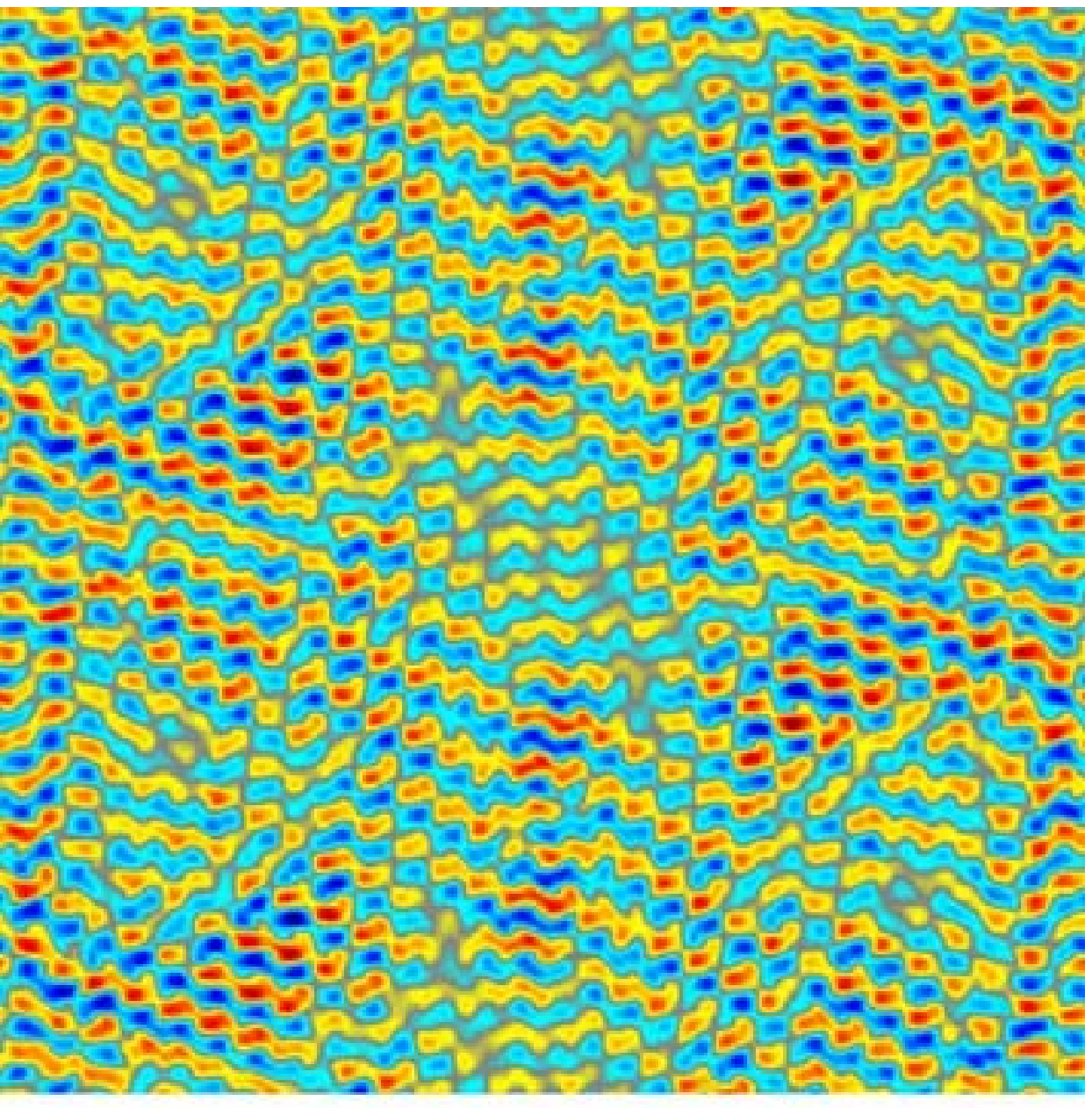}
    \hspace{0mm}
    \includegraphics[width=0.45\textwidth]{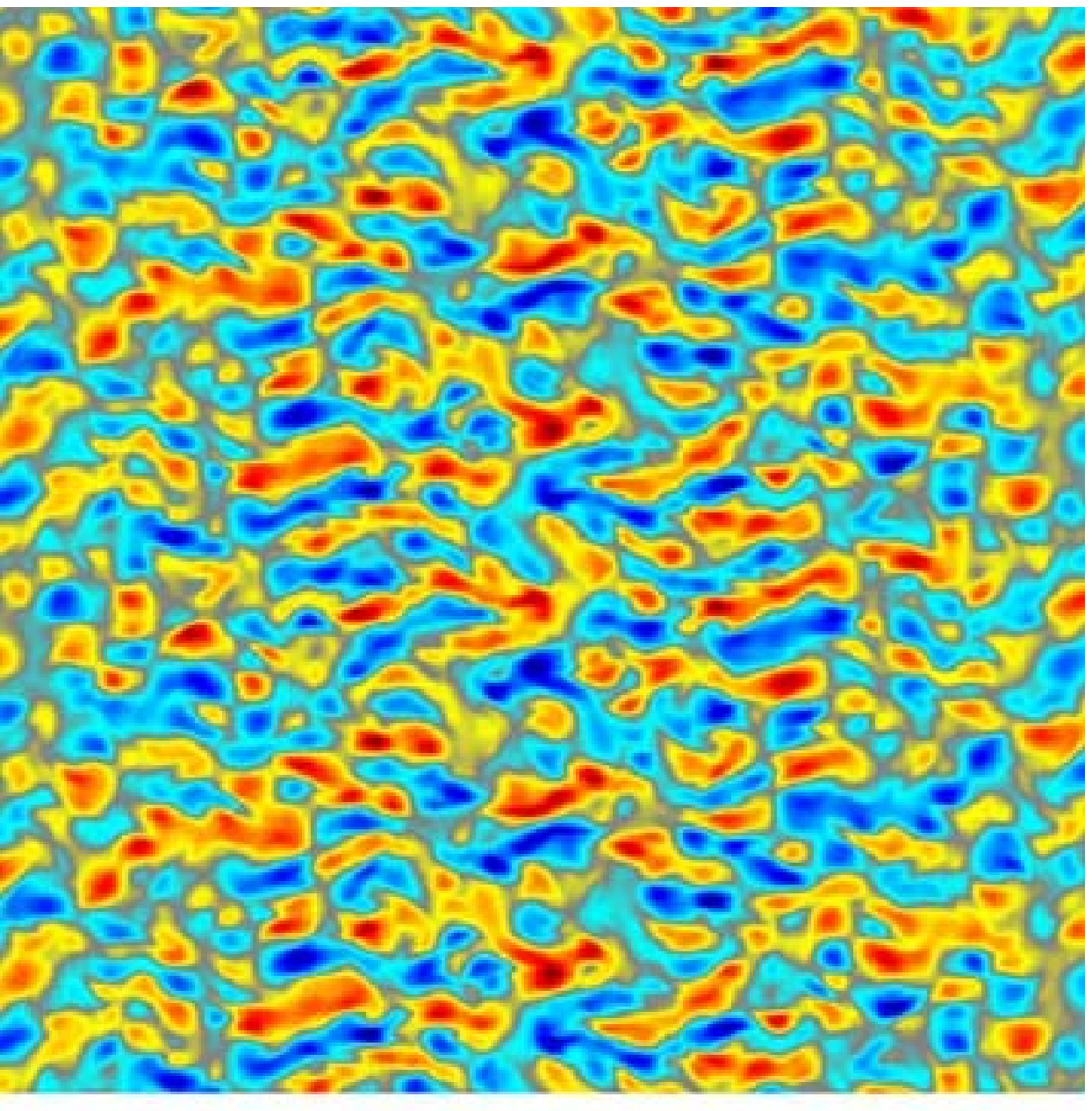}
    \\[0pt]
    \caption{KH instability in a vortical plane wave.
    Snapshots of the $x$-component of the velocity are shown. Parameters are
    $N=512$, $k_{x,0}/k_y=-16$, and $\delta v_{x,0}/c_s=10^{-2}$.  Images are
    at intervals of $0.5\Omega^{-1}$ from the beginning of the run. The planar
    shearing wave is seriously distorted after only 1 shear time ($\Omega t=1$).}
    \label{figKH}
\end{figure*}

Since with sufficient resolution it is possible to accurately
follow the amplification of incompressible waves by factors of
$10^{3-4}$, it is of interest to investigate what occurs when the
wave reaches nonlinear amplitude ($\delta v_x > c_s$) at peak
amplification. Interestingly, we have found that for some initial
conditions incompressible plane wave solutions become unstable to
Kelvin-Helmholtz modes quickly.
% as they approach nonlinear amplitudes.
% I have removed the last half of the sentence, since it gives some
% confusion that the transient amplification helps destabilize the flow. (Yue)
Figure \ref{figKH} shows the development of the KH instability in
shearing wave with $k_y=4\pi/L_y$, $k_{x,0}/k_y=-16$, and initial
amplitude $\delta v_{x,0}/c_s=10^{-2}$ on a $512^2$ grid. Four
images of $\delta v_x$ are shown at $\Omega t = 0, 0.5, 1.0$ and
1.5. The destruction of the plane wave is clearly evident. We have
confirmed that the presence of the instability is independent of
the numerical resolution. Runs on a $1024^2$ grid show the same
behavior, although since the modes are seeded by grid noise in the
initial conditions, the details change at higher resolution.

\begin{figure*} \centering
\includegraphics[width=0.45\textwidth]{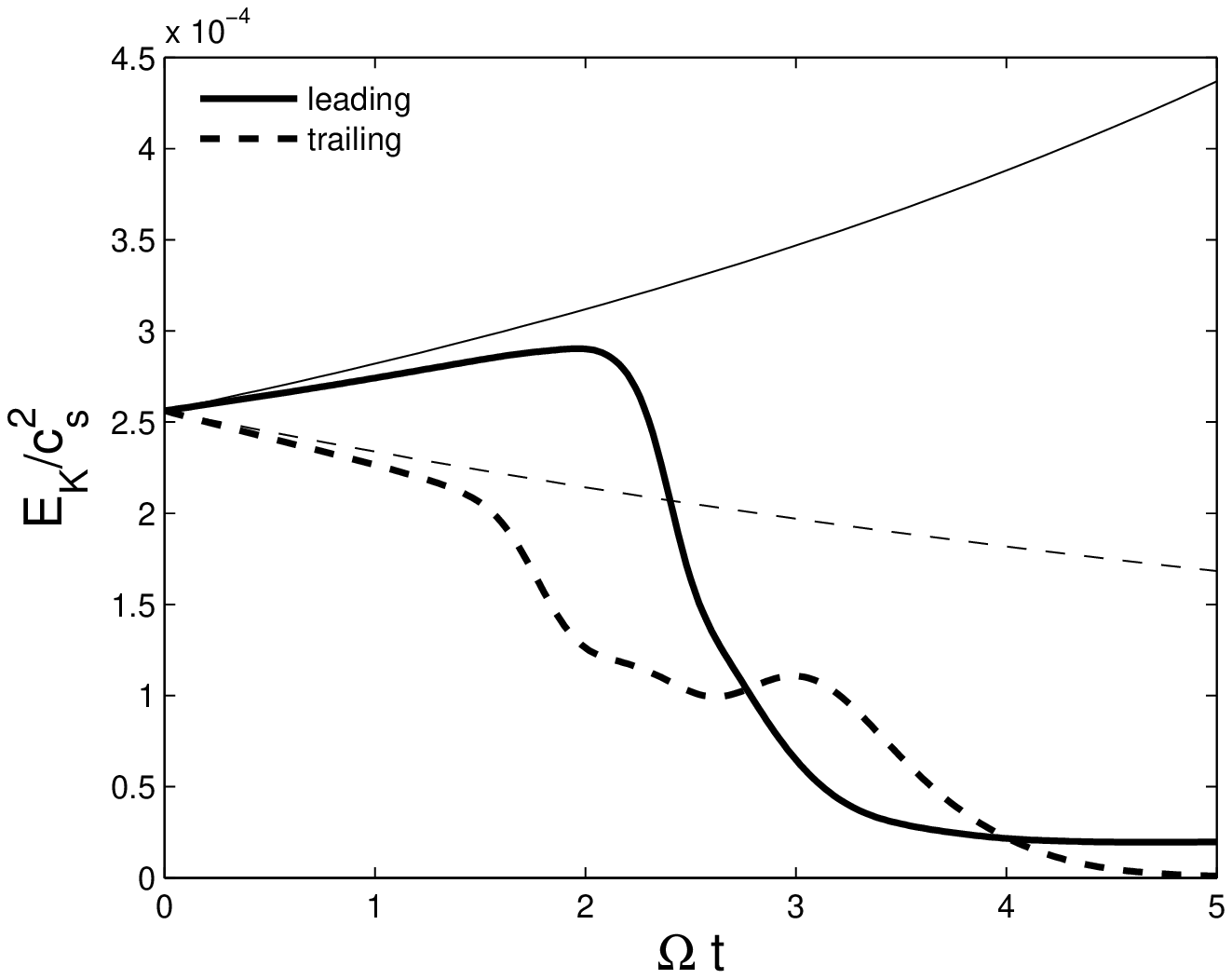}
\includegraphics[width=0.45\textwidth]{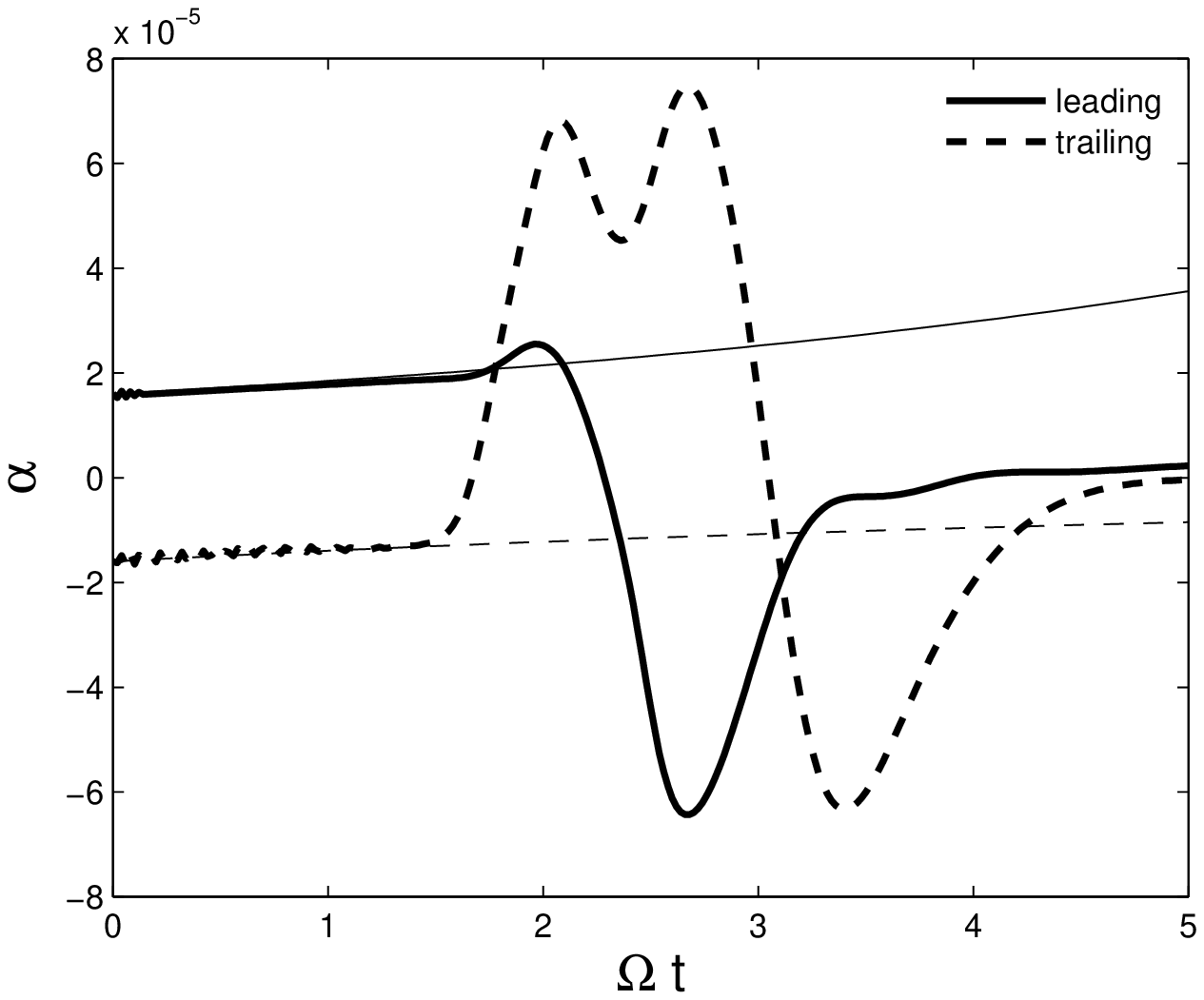}
\caption{Evolution of volume-averaged {\em left:} kinetic energy,
and {\em right:} shear stress normalized by the gas pressure in KH
unstable leading and trailing incompressible plane waves. The
parameters are $\delta v_{x,0}/c_s=10^{-3}$, $k_y=4\pi/L_y$,
$k_{x,0}/k_y=\mp 32$ and $N=1024$. Light curves are expected
analytic solutions and heavy curves are numerical results.}
\label{figKHhistory}
\end{figure*}

Since the KH instability depends only on the amplitude of the
shear but not the sign, we should expect it to affect both leading
and trailing waves.  By repeating the evolution shown in figure
\ref{figKH} but with $k_{x,0}/k_y = +16$, we have confirmed that
even though the wave amplitude is now decaying, it is still KH
unstable. In fact, the onset of the KH instability is not related
to the transient amplification from leading to trailing, but is
primarily determined by the initial condition. We now give some
detailed analysis below.

It is instructive to use the KH instability criterion in planar
flows to interpret the evolution of plane waves in the shearing
sheet, even though this criterion ignores the effect of the
Coriolis and tidal gravity forces. In a uniform density medium, a
shear profile with an inflection point is always unstable, with a
growth rate $n \sim k \delta v$.  For shearing vortical waves, we
might expect instability if the KH growth rate exceeds the
angular velocity, that is
\begin{equation}\label{KHcond}
\mid k_x \delta v_y\mid \gtrsim \Omega
\end{equation}
In fact, this crude criterion seems to apply remarkably well to
our results. For example, simulations with
%$\mid k_{x,0}\mid\ge80\pi$ and $\delta v_{x,0}=10^{-5}$ are unstable,
%while those with $\mid k_{x,0}\mid\le32\pi$ and $\delta
%v_{x,0}=10^{-5}$ are stable.
% Jim, I have used two different examples because I am now using the trailing
% phase (instead of the leading phase) to judge the stability of solution; see
% my discussions below.
$\mid k_{x,0}\mid\ge128\pi$ and $\delta v_{x,0}=10^{-6}$ are
unstable, while those with $\mid k_{x,0}\mid\le32\pi$ and $\delta
v_{x,0}=10^{-6}$ are stable (note $k_y\equiv 2(2\pi/L_y) =
8\pi$ in these tests, and $\mid \delta
v_{y,0}/\delta v_{x,0}\mid = \mid k_{x,0}/k_y\mid$).
By ``stable'' we mean the shearing wave solution survives in
both the leading phase and the trailing phase.

However, the KH instability of plane shearing waves is not related
to shear amplification. Note that $\mid k_x\delta v_y\mid$ in
(\ref{KHcond}) is not constant, but evolves with time. The
evolution of $\mid k_x\delta v_y\mid$ can be described by (if the
incompressible plane wave solution is stable):
\begin{equation}\label{KH_evo}
\mid k_x\delta v_y\mid=\mid\delta v_{x,0}k_y\mid{\cal
A}\bigg[1-\frac{1}{1+(k_{x,0}/k_y+q\Omega t)^2}\bigg]\ .
\end{equation}
Thus $\mid k_x\delta v_y\mid$ starts from the initial value $\mid
\delta v_{x,0}k_y\mid ({\cal A}-1)$, decreases in the leading
phase and becomes zero at $t_{\rm max}$; after that, it increases
again in the trailing phase and approaches the limit of $\mid
\delta v_{x,0}k_y\mid {\cal A}$. Therefore although during the
shear amplification process, the perturbation amplitude $\delta
v_x$ is amplified, the instability condition is not violated. In
fact, the derivation of criterion (\ref{KHcond}) only considers
azimuthal velocity shear. If one further includes radial shear,
criterion (\ref{KHcond}) will be modified. The KH instability
growth rate is directly related to the local maximum of vorticity
(e.g., Drazin \& Reid 1981), i.e., $n\sim \mid W_{\rm
max}\mid\equiv\mid k_x\delta v_y - k_y\delta v_x\mid \equiv
\mid\delta v_{x,0}k_y\mid {\cal A}$, the latter equal sign
according to the analytical solution (\ref{dVx}). Therefore
criterion (\ref{KHcond}) is modified into
\begin{equation}\label{KHcond_mod}
\mid W_{\rm max}\mid\equiv\mid\delta v_{x,0}k_y\mid {\cal
A}\gtrsim \Omega\ .
\end{equation}
Hence the KH instability is determined by the initial conditions.
However, due to the subtlety that the shear is not steady and the
position of maximum vorticity is changing, the trailing phase
might have a slight preference of KH instability because $\mid
k_x\mid$ increases during the trailing phase and the wave becomes
more and more quasi-steady.

%Due to the fact that the shear is not steady and $\mid k_x\delta
%v_y\mid$ is evolving, the wave could be stabilized during the
%leading to trailing process even if the initial conditions
%satisfies criterion (\ref{KHcond}), but it will eventually become
%unstable during the trailing phase.

Figure \ref{figKHhistory} ({\em left}) shows the evolution of the
volume averaged kinetic energy in both leading and trailing waves
with parameters $\delta v_{x,0}/c_s=10^{-3}$, $k_y=4\pi/L_y$,
$k_{x,0}/k_y=\mp 32$ on a $1024^2$ grid along with the expectation
from linear theory. For the parameters adopted above, peak
amplification should occur at $\Omega t = 21.3$ for the trailing
case. For both leading and trailing waves, the numerical solution
begins to diverge significantly from the analytic curve after
$\Omega t \sim 2$ of evolution. The kinetic energy rapidly decays
after the KH instability saturates.  The linear growth rate of the
KH instability can be computed from the time evolution of the
amplitude of the numerical versus analytic plane-wave solution; we
measure a growth rate of about $0.2 \mid W_{\rm max}\mid$,
%where $W_{\rm max}$ is the maximum of vorticity,
as expected for
the fastest growing mode (Drazin \& Reid 1981).

Figure \ref{figKHhistory} ({\em right}) shows the evolution of the
volume averaged angular momentum transport $\alpha$ in both
leading and trailing waves with the same parameters as in Figure
\ref{figKHhistory}a.  The KH instability causes the value of
$\alpha$ to (1) change sign, and (2) increase by a factor of about
two compared to the expectation of linear theory. Note however
that at late times $\alpha$ drops to near zero, that is there is
no evidence for sustained transport.  Thus, although it might
appear attractive to suggest the KH instability is a route to
turbulence and sustained transport, our simulations do not support
this idea. We discuss this further in \S 5.

We suggest that the reason previous studies (Umurhan \& Regev
2004; JGb) did not find destruction of shearing vortical waves is
that they used only small initial perturbation amplitudes and
intermediate amplification factors (small $\mid k_{x,0}/k_y\mid$),
which did not satisfy the KH instability condition equation
(\ref{KHcond_mod}).
%lead to wave amplitudes that satisfy the KH instability condition equation
%(\ref{KHcond}).
% I modified the last sentence (Yue).

\section{EVOLUTION OF VORTICES IN TWO- AND THREE-DIMENSIONS}

\begin{figure*}
  \centering
    \includegraphics[width=0.45\textwidth]{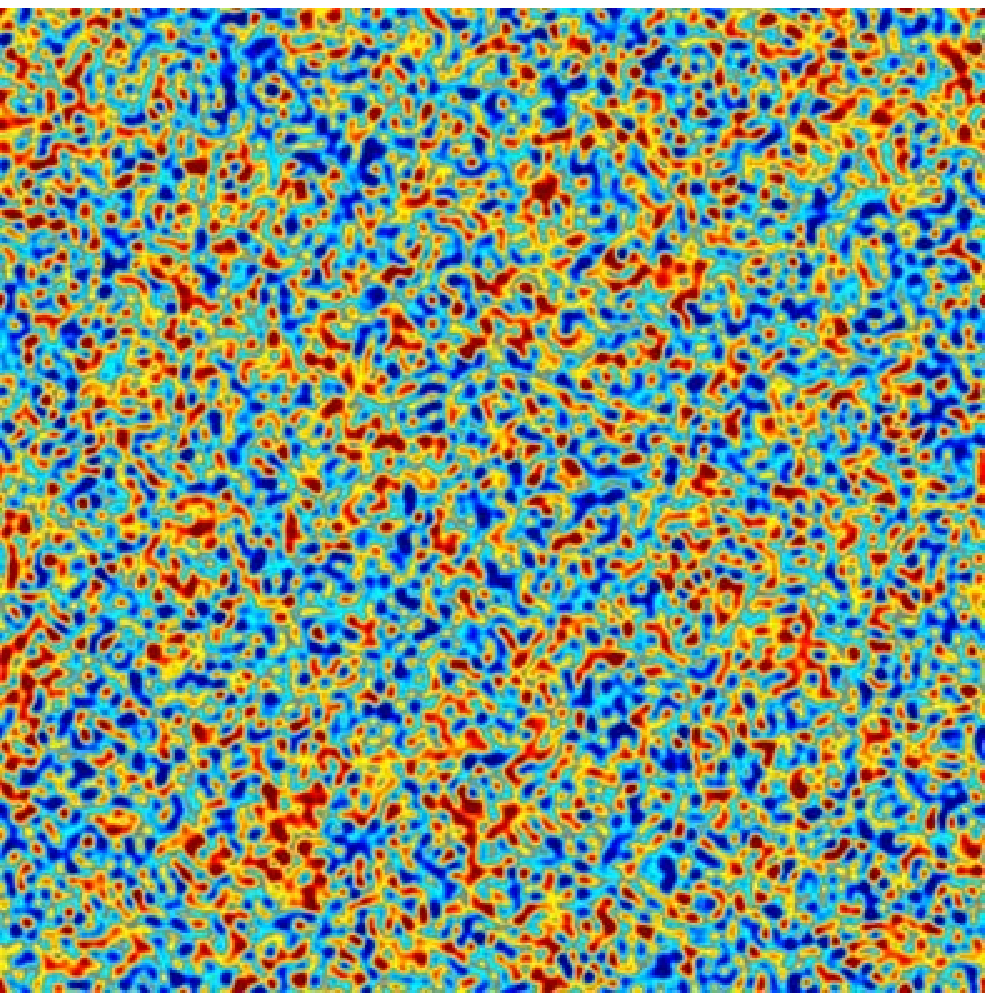}
    \hspace{0mm}
    \includegraphics[width=0.45\textwidth]{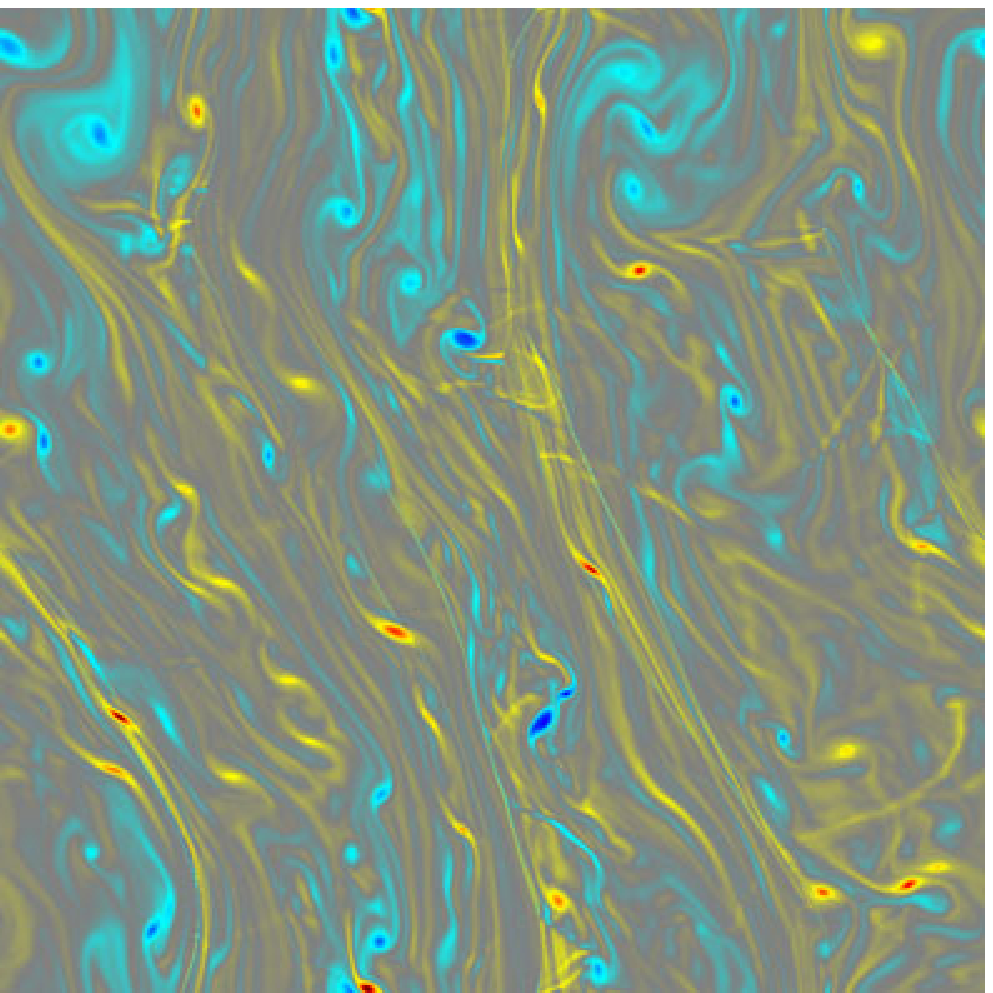}
    \\[6pt]
    \includegraphics[width=0.45\textwidth]{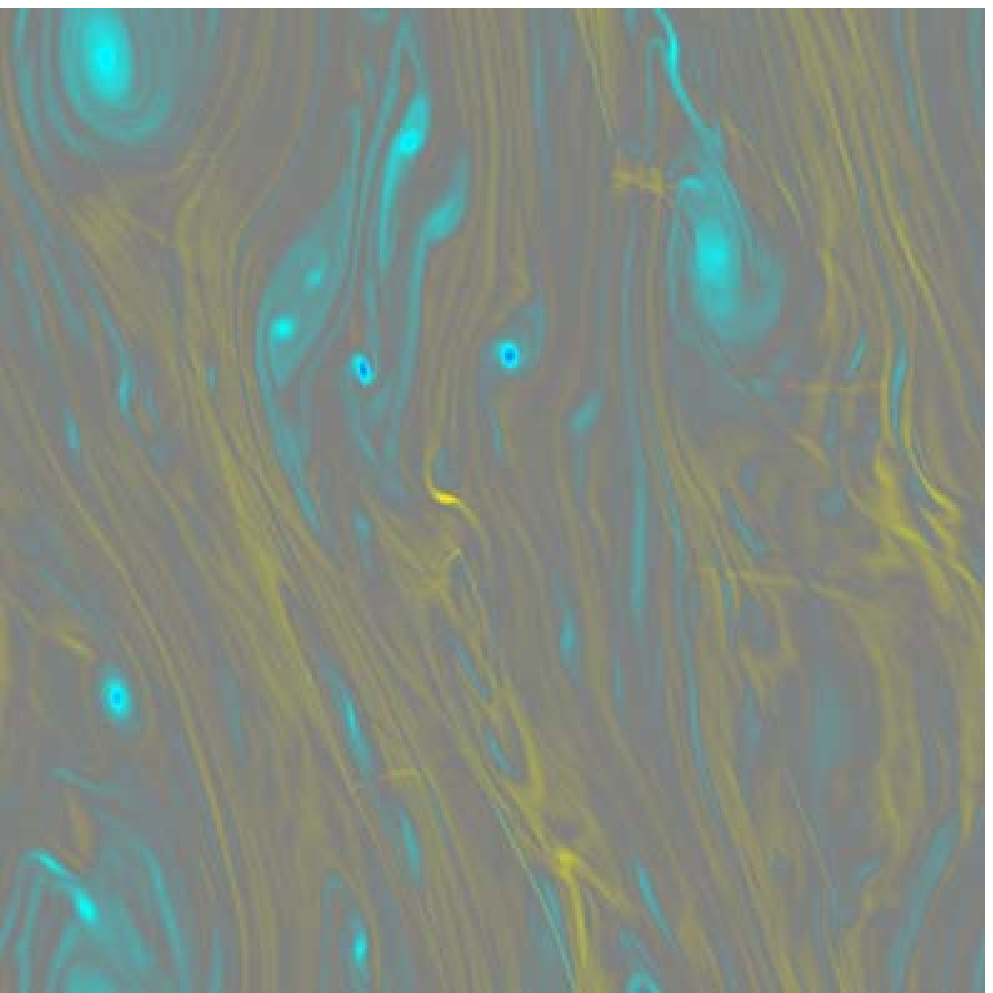}
    \hspace{0mm}
    \includegraphics[width=0.45\textwidth]{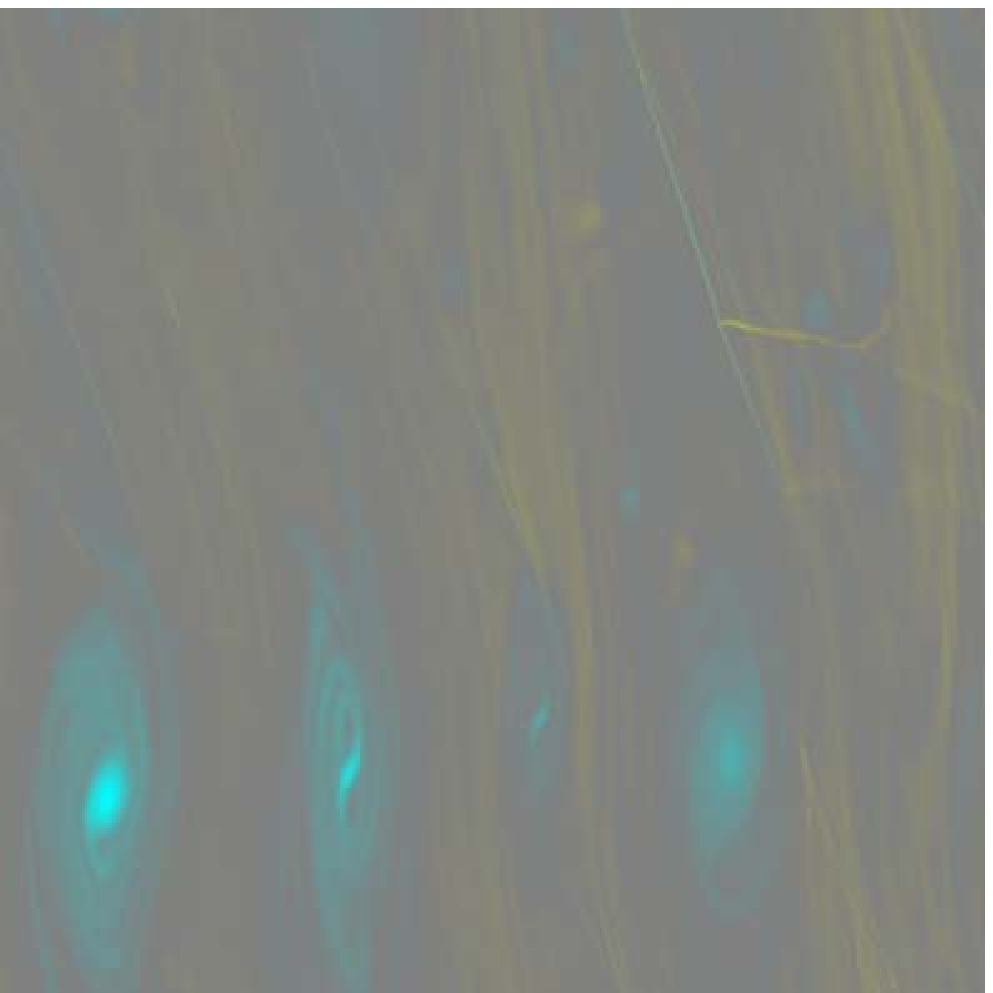}
%    \includegraphics[width=80mm]{HSB.whole.0002.d.eps}
%    \hspace{0mm}
%    \includegraphics[width=80mm]{HSB.whole.0006.d.eps}
    \\
    \caption{Evolution of the perturbed $z$-component of vorticity $\delta W_z
= W_z+q\Omega$ in 2D starting from a random distribution with a
power spectrum consistent with Kolmogorov at a resolution of
$1024^2$.  The box size is
    $4H\times 4H$. The initial state is shown in the upper left panel. Frames are
    taken in lexicographic order at $\Omega t=0,\ 10,\ 20,\ 60$. Only anticyclonic vortices (blue)
    survive at late time.}
    \label{fig2DRan}
\end{figure*}

\begin{figure*}
  \centering
    \includegraphics[width=0.45\textwidth]{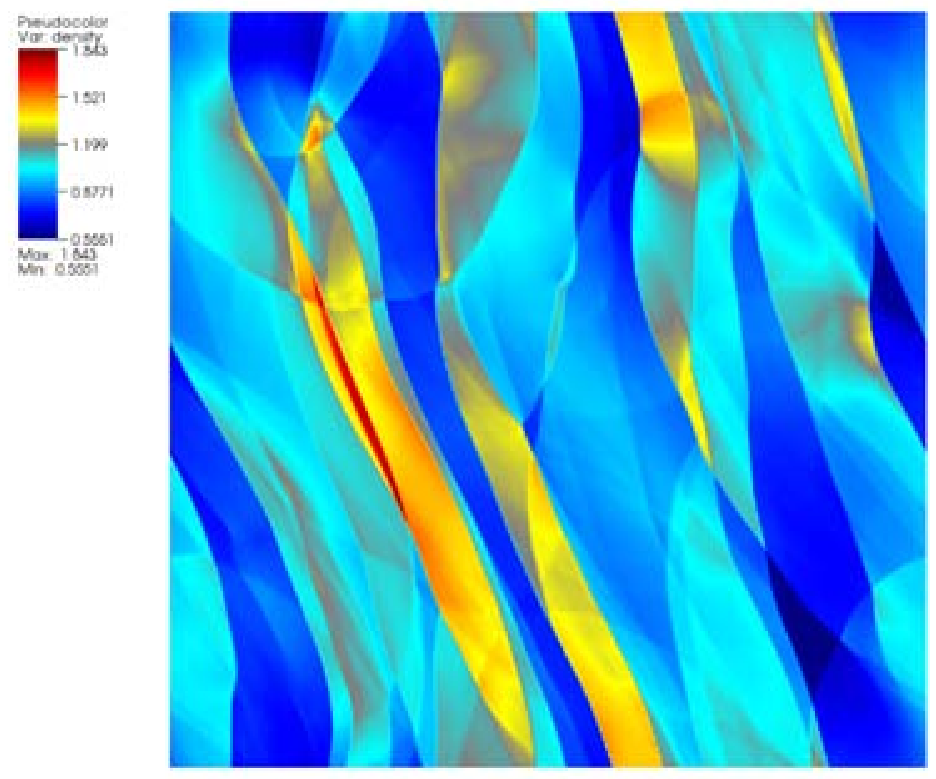}
    \includegraphics[width=0.45\textwidth]{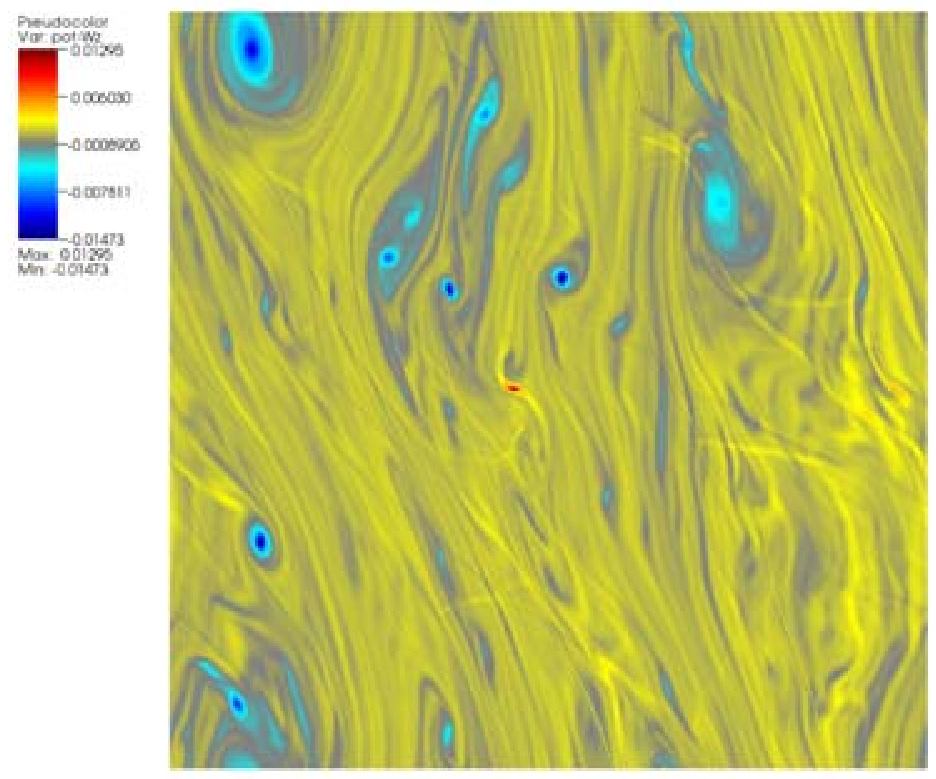}
    \caption{Snapshots of the density ({\em left}) and potential vorticity
$\tilde{W}_z = (W_z+2\Omega)/\rho$ ({\em right}) at $\Omega t=20$
in the 2D random $1024^2$ run.}
    \label{2DRan_Den_Wz}
\end{figure*}

We now consider the evolution of random vortical (incompressive)
velocity perturbations in both two- and three-dimensions.  Our
study closely parallels the work of JGb. We draw the initial
velocity perturbations from a Gaussian random field with a 2D
Kolmogorov power spectrum $\mid\delta {\bf v}\mid^2\sim k^{-8/3}$.
We apply a cut-off to the spectrum at $k_{\rm min}=\pi/H$ and
$k_{\rm max}=32k_{{\rm min}}$.  The amplitude of the random
perturbations is characterized by $\sigma=\langle\mid\delta {\bf
v}\mid ^2\rangle^{1/2}$, we use $\sigma/c_s=0.8$ throughout. The
effect of varying $\sigma$ was explored in JGb; smaller values
lead to less angular momentum flux (see below).  To ensure the
perturbations are initially incompressible, we actually compute
the $z-$component of a
velocity potential $A_z$ with the appropriate power
spectrum, and then compute the velocity perturbations from ${\bf
v} = \nabla \times {\bf A}$.  To study the effect of numerical
resolution, we have found it critical to compute the power
spectrum in Fourier space once and for all, and then use this same
spectrum to generate the initial conditions at every resolution.
This is only possible because we cut-off the power spectrum at
$k=k_{\rm max}$, so there is zero power in the additional high-$k$
modes that can be represented with higher spatial resolution.

As before, the simulations all use $\rho_0=1$, $c_s=10^{-3}$, $\Omega=10^{-3}$
and a box size of $L_x=L_y=4H$ in 2D (and $L_z=H$ in 3D).

\subsection{Random Vorticity in Two-Dimensions}

We have run simulations with resolutions from $N=128$ up to
$N=2048$. Figure \ref{fig2DRan} shows snapshots in the evolution
of the vorticity fluctuations $\delta{W}_z = W_z+q\Omega$ for the
$1024^2$ case. As in previous work, we confirm that cyclonic
vortices (rotating in the same sense as background shear) with
positive values of $\delta {W}_z$ are quickly destroyed, while
anticyclonic vortices with negative value of $\delta {W}_z$
survive and decay slowly.  This results in large vortices with
negative $\delta {W}_z$ embedded in a smooth background with a
slightly positive $\delta {W}_z$.  The volume average of the
vorticity fluctuations reduces to a line integral along the
boundaries of our domain.  Since we use periodic boundaries in
$y$, this integral is a measure of the accuracy of shearing sheet
boundary conditions. We find $\langle \delta {W}_z \rangle \sim
10^{-12}$, i.e. essentially zero, as expected.

Compressibility does play an important role in the dynamics; figure
\ref{2DRan_Den_Wz} is a snapshot of the density and potential vorticity
$\tilde{W}_z$ at $\Omega t = 20$ for the $N=1024$ simulation.  Strong,
interacting shocks which propagating primarily in the radial direction are
clearly evident in the image of the density.  The density contrast is more
than a factor of three between the lowest and highest density regions.
The locations of density jumps associated with shocks is not immediately
evident in the plot of $\tilde{W}_z$.  However, horizontal streaks of positive
$\tilde{W}_z$ are visible, and are associated with shock intersections.
As the shocks propagate over vortices, they are refracted, producing
curvature in the shock front visible in the density image.  At the same
time, the vortices are impulsively accelerated at each shock passage.
Since shocks propagate in both radial directions, vortices do not gain
any net radial motion from shock accelerations, but are perturbed back
and forth.

\begin{figure} \centering
\includegraphics[width=0.5\textwidth]{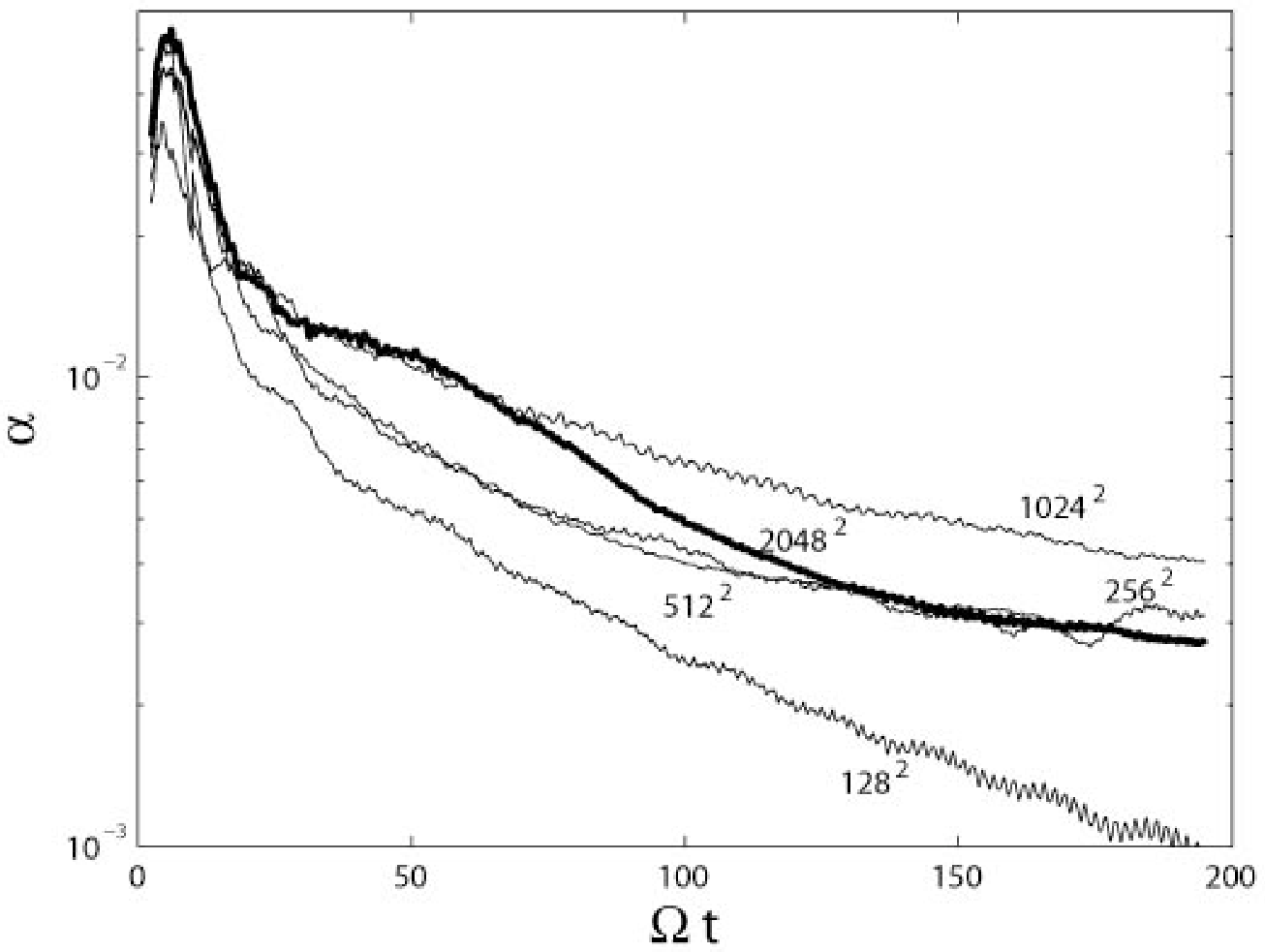}
\includegraphics[width=0.5\textwidth]{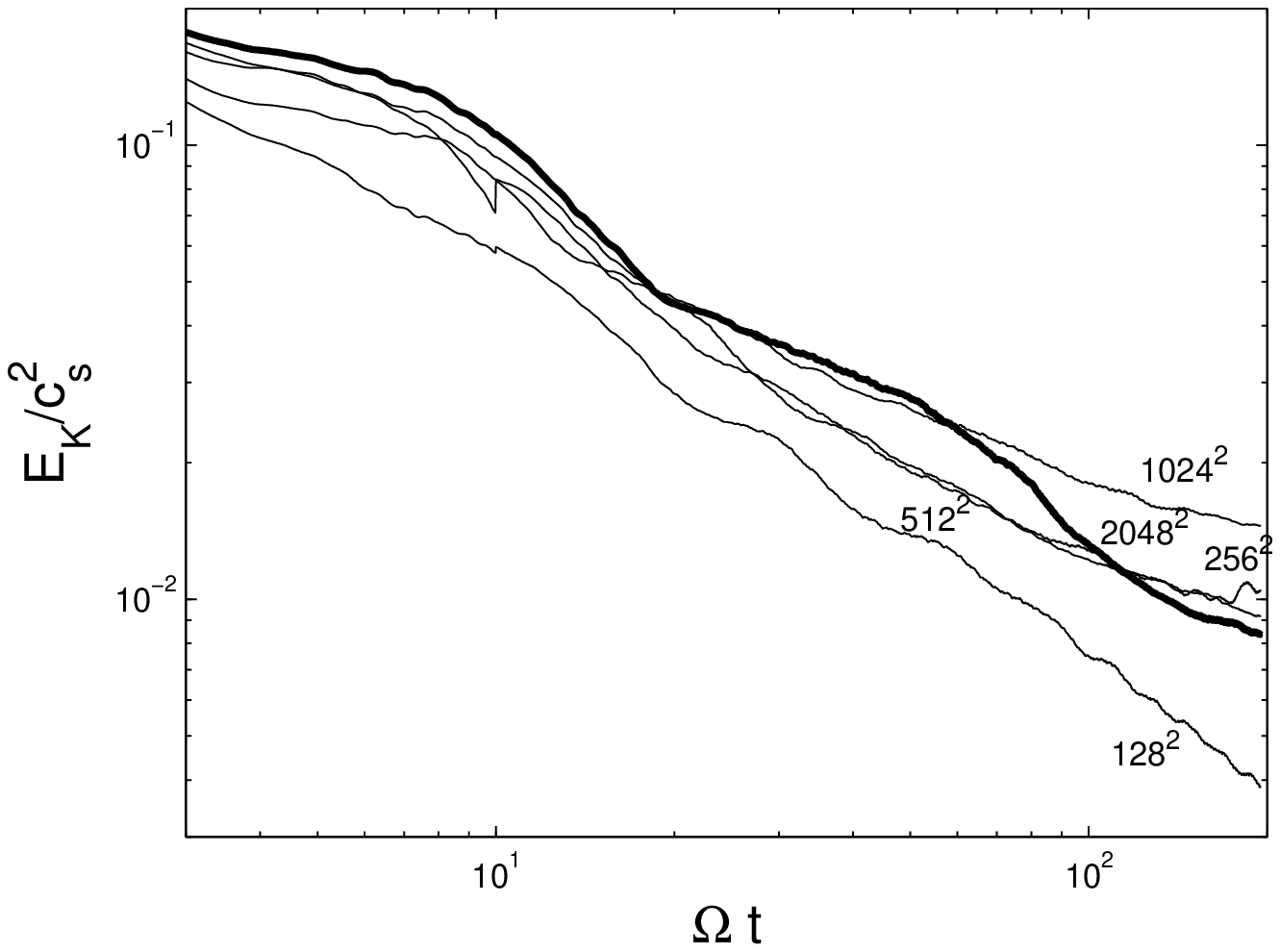}
\caption{Time evolution of the volume-averaged dimensionless shear
stress $\alpha$ ({\em upper}) and 2-dimensional kinetic energy
density ({\em bottom}) for the 2D random vorticity runs.}
\label{2Drandom}
\end{figure}

We plot the time evolution of $\alpha$ and $E_{{\rm K}}$ at each
resolution in figure \ref{2Drandom}. The data points have been
boxcar smoothed over an interval $\Delta t=10\Omega^{-1}$ (a
smaller boxcar size $\Omega\Delta t=5$ is used for data points
prior to $\Omega t=10$). Compared with fig. 4 in JGb, our $128^2$,
$256^2$ and $512^2$ runs closely resemble their $256^2$, $512^2$
and $1024^2$ runs, and show some ``convergence'' towards higher
resolution. However, our higher resolution $1024^2$ and $2048^2$
runs do not follow this trend.  The time-averaged value of
$\alpha$ for the last half of the runs ($100<\Omega t<200$) is of
order $\sim 10^{-3}$, consistent with the highest resolution runs
in JGb.  Both $\alpha$ and $E_{K}$ decay by more than an order of
magnitude during the evolution, and this decay shows no signs of
ending.  We find there can be significant differences at late
times between simulations that use the same initial power-law
spectra, but different initial amplitudes and phases for
individual Fourier modes. We attribute this to statistical
fluctuations in the small number of small $k$ modes between
different realizations of the same power spectrum. The evolution
of small $k$ modes depends on mode interactions at large
amplitudes, and the numerical dissipation rate at linear
amplitudes. Since the numerical dissipation of features resolved
by 16 grid points or larger is negligible in our methods, once
these small $k$ modes reach linear amplitude they decay extremely
slowly, and can dominate the late-time evolution.  Thus, the
pattern of large-scale modes that emerges and decays at late times
will depend on the initial conditions, and the history of mode
interactions that occur during the evolution. Either way, because
of the small number of small $k$ modes available, we expect
significant statistical fluctuations between different runs at
late times.  Our use of boxcar averaging to smooth the data has
eliminated the high-frequency fluctuations that dominate both
$\alpha$ and $E_{K}$ at late times, with typical amplitudes about
twice the averaged values in the plots.

The lack of convergence of $\alpha$ and $E_{{\rm K}}$ in the
$1024^2$ and $2048^2$ runs bears some discussion.  Note that at
early times, both show the same amplitude at $\Omega t =50$, but
at a much higher level than all the lower resolution runs.
Thereafter, the $2048^2$ run decays to the same level as the other
runs whereas the $1024^2$ run does
not. This evolution seems to be related to the precise
distribution of large-scale vortices that emerge in the flow.  At
$\Omega t=50$, images of the vorticity show four strong,
well-defined vortices at the higher resolutions, whereas at lower
resolution only 2-3 weaker and more diffuse vortices are evident.
There are also a significant number of small, weak vortices at higher
resolution.
Thus, the differing rates of diffusion and merging of vortices at
different resolutions may help to explain the histories.

\begin{figure*}
  \centering
    \includegraphics[width=0.45\textwidth]{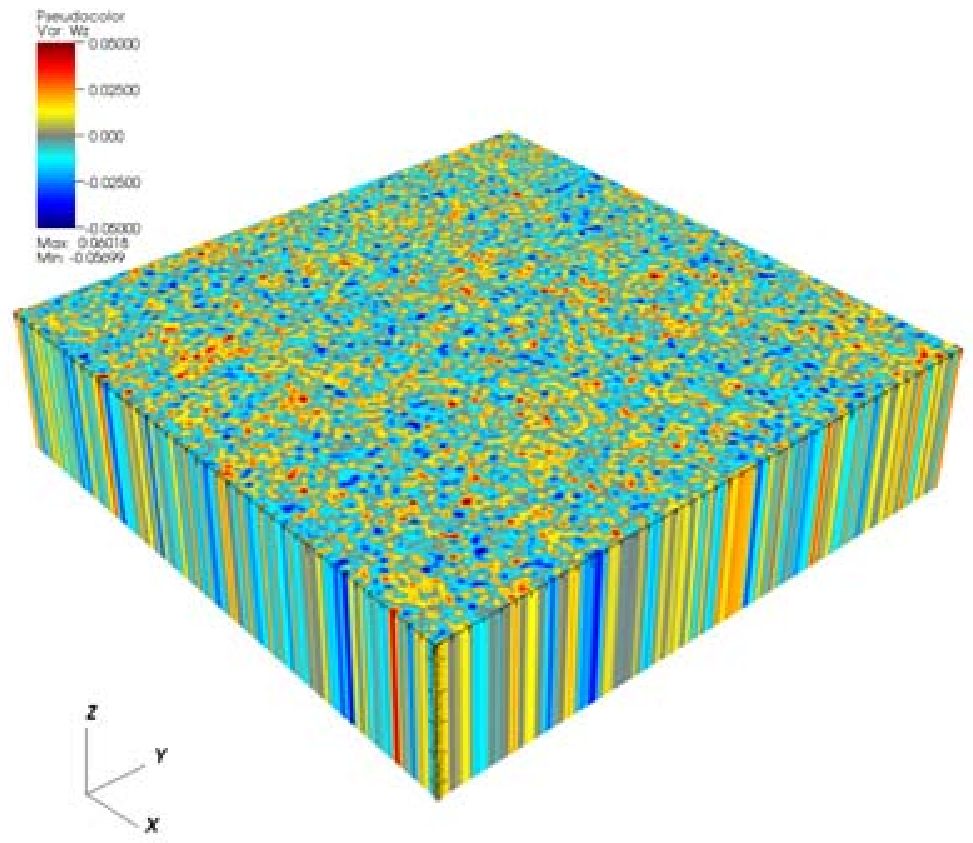}
    \includegraphics[width=0.45\textwidth]{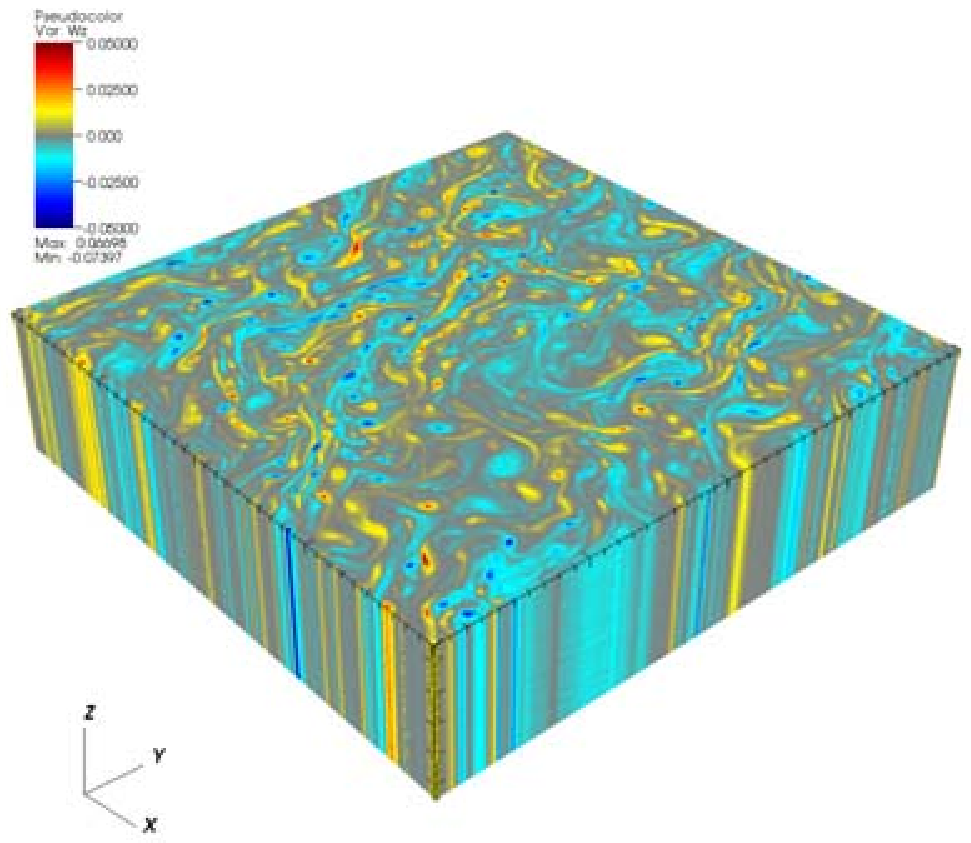}
    \\[-10pt]
    \includegraphics[width=0.45\textwidth]{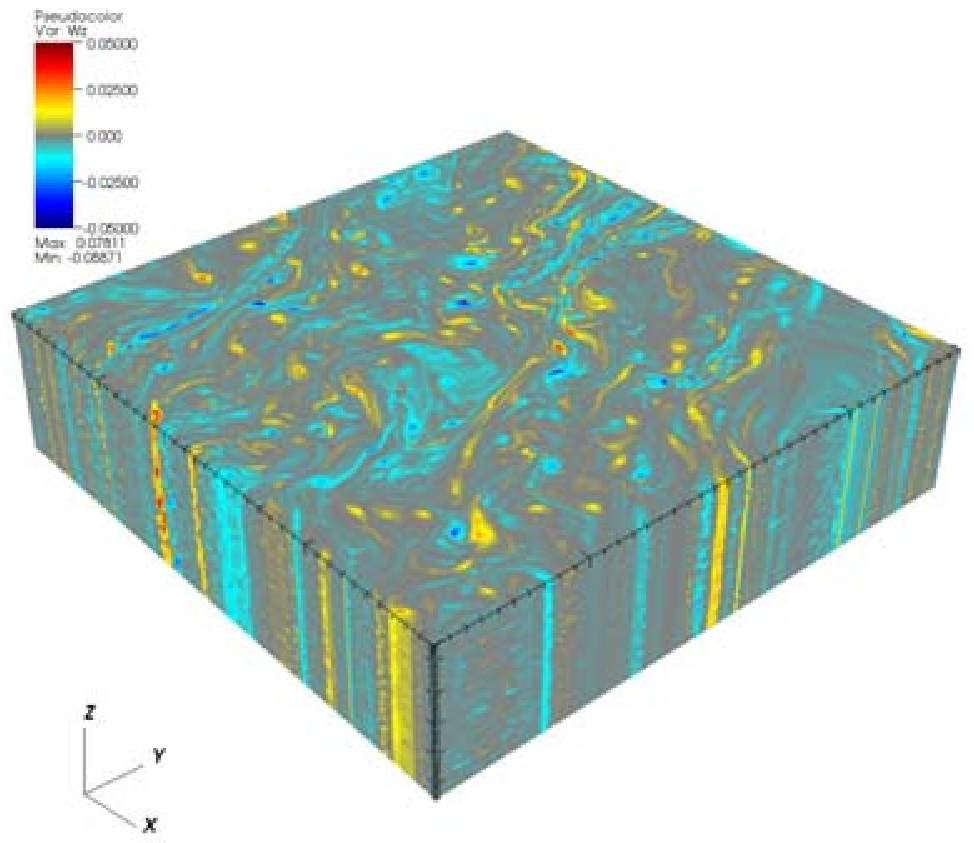}
    \includegraphics[width=0.45\textwidth]{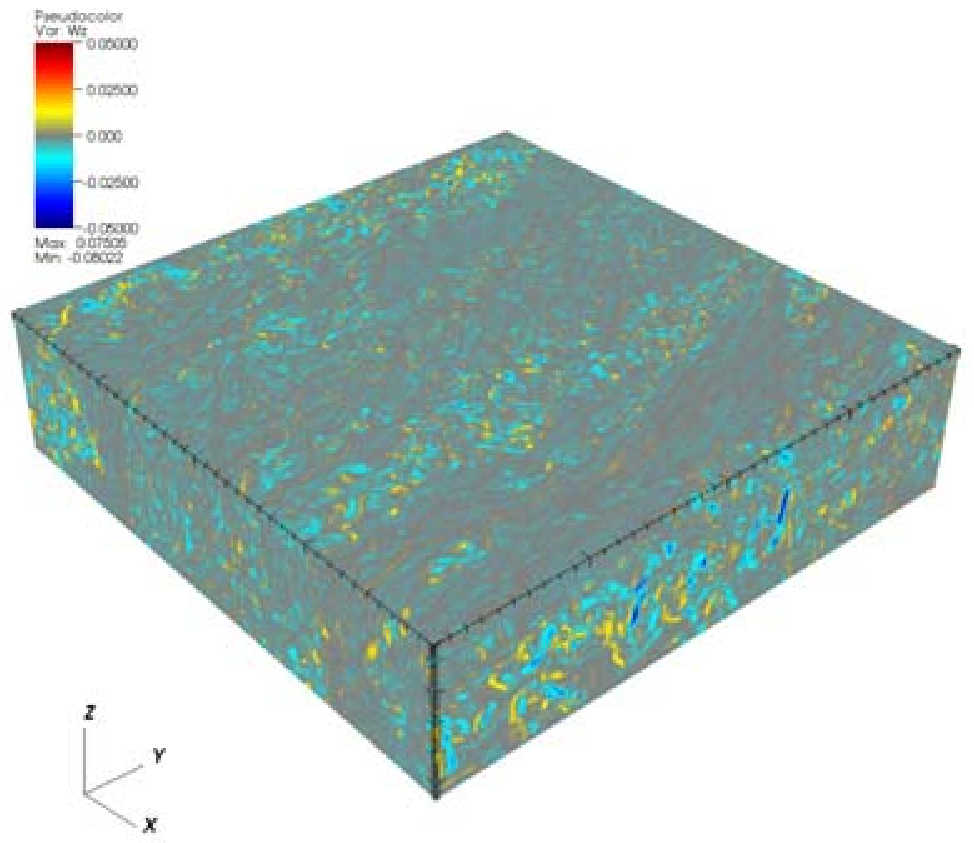}
    \\[-10pt]
    \caption{Slices of the $z$ component of vorticity $W_z$ in the $512^2\times 128$ 3D random
    vorticity run. Snapshots are taken at $\Omega t=0,\ 10,\ 20,\ 60$.}
    \label{fig3DRan}
\end{figure*}

\begin{figure} \centering
\includegraphics[width=0.5\textwidth]{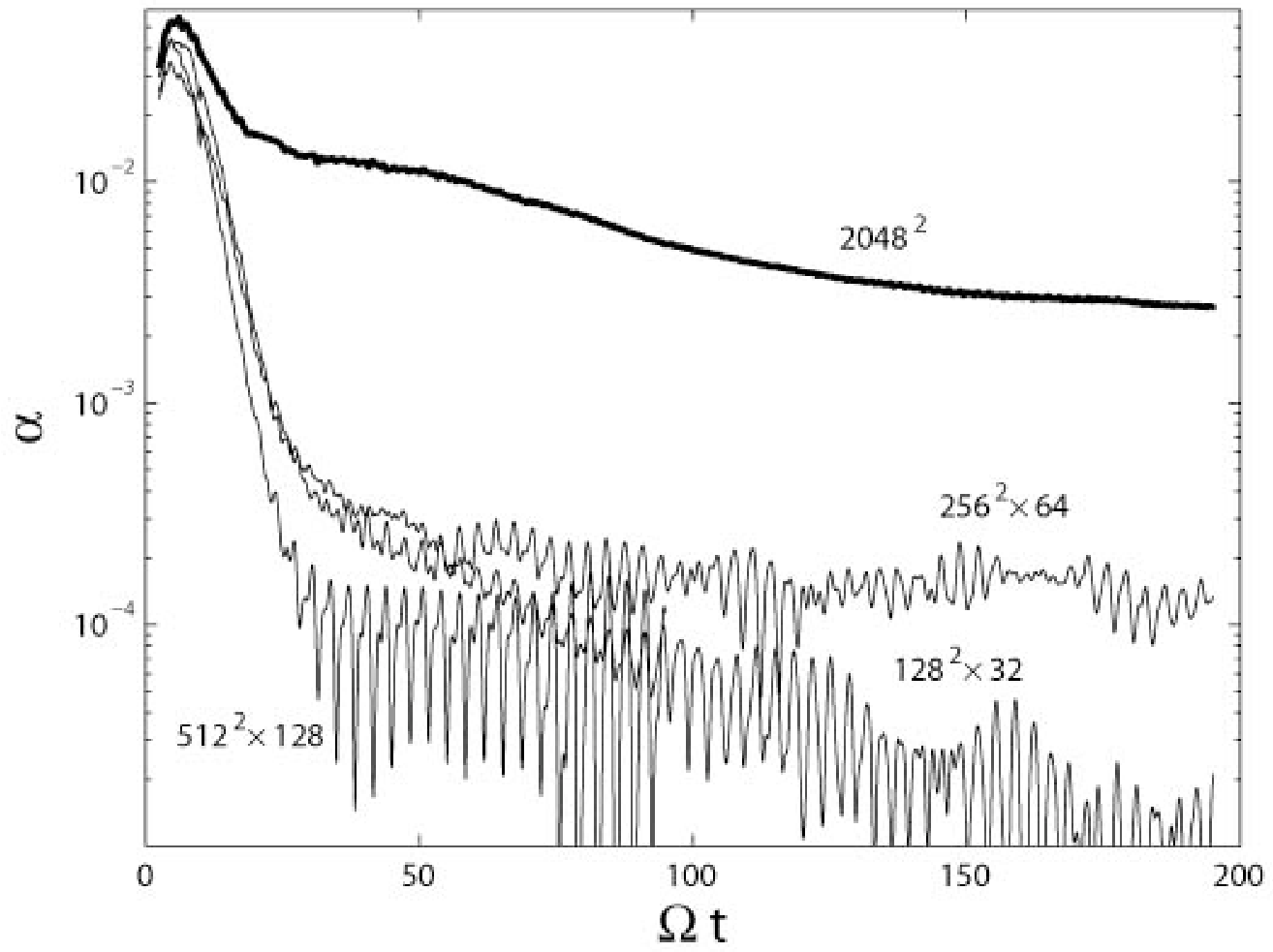}
\includegraphics[width=0.5\textwidth]{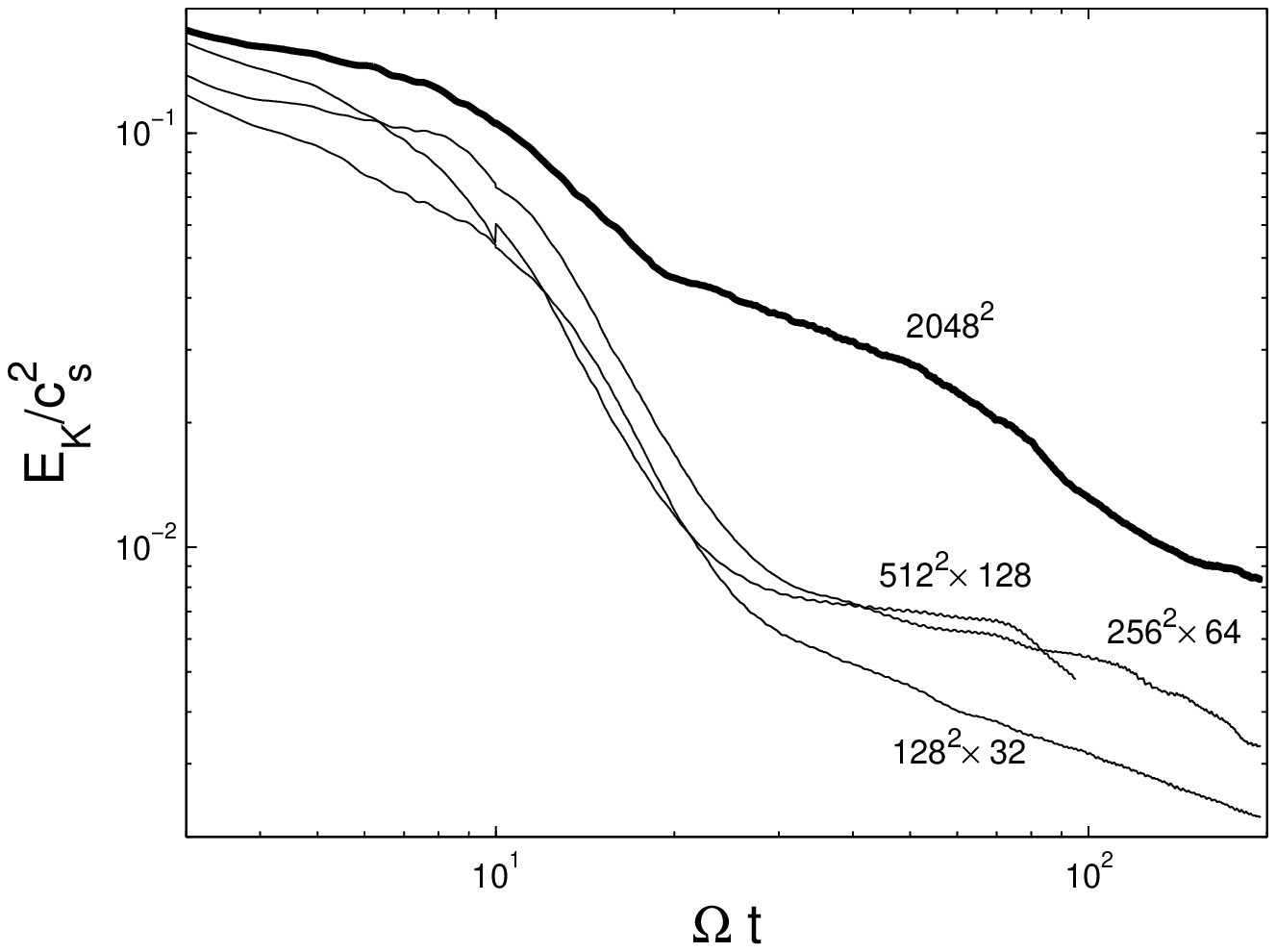}
\caption{Time evolution of the volume-averaged dimensionless shear
stress $\alpha$ ({\em upper}) and 2-dimensional kinetic energy
density ({\em bottom}) for the 3D random vorticity runs. The heavy
solid curve is the result of the $2048^2$ 2D run for reference.}
\label{3Drandom}
\end{figure}

\begin{figure*}
  \centering
    \includegraphics[width=0.45\textwidth]{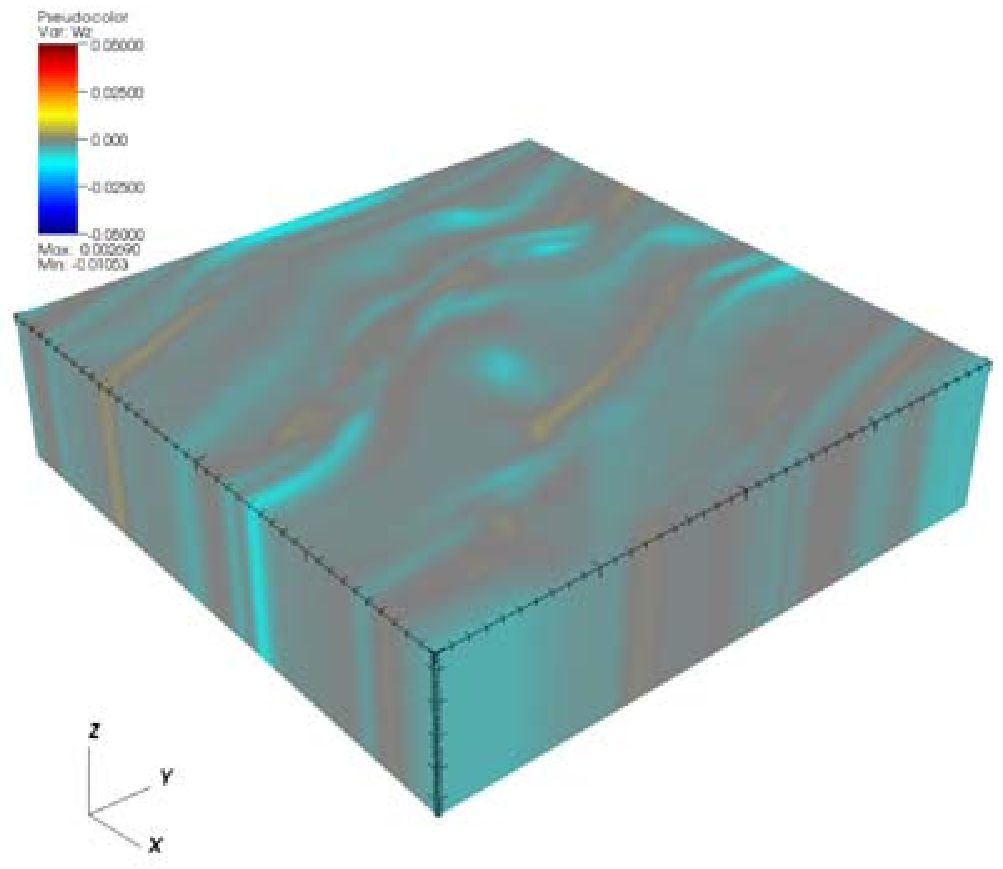}
    \includegraphics[width=0.45\textwidth]{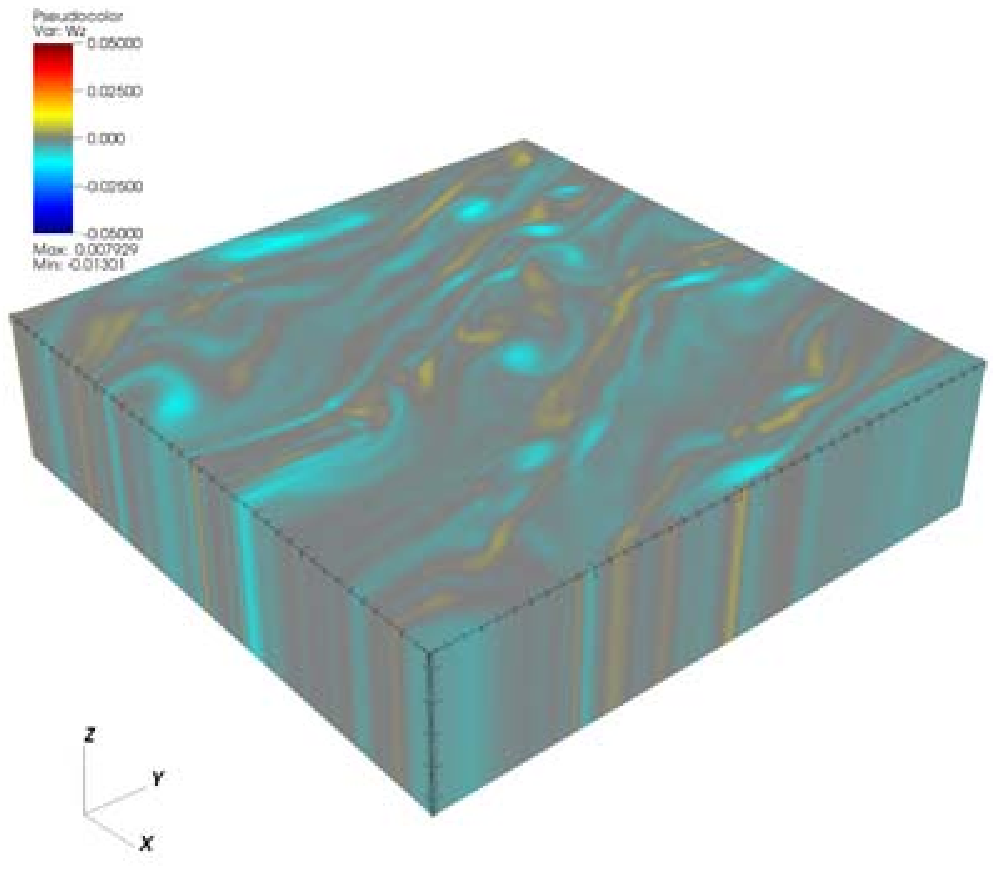}
    \\[-10pt]
    \includegraphics[width=0.45\textwidth]{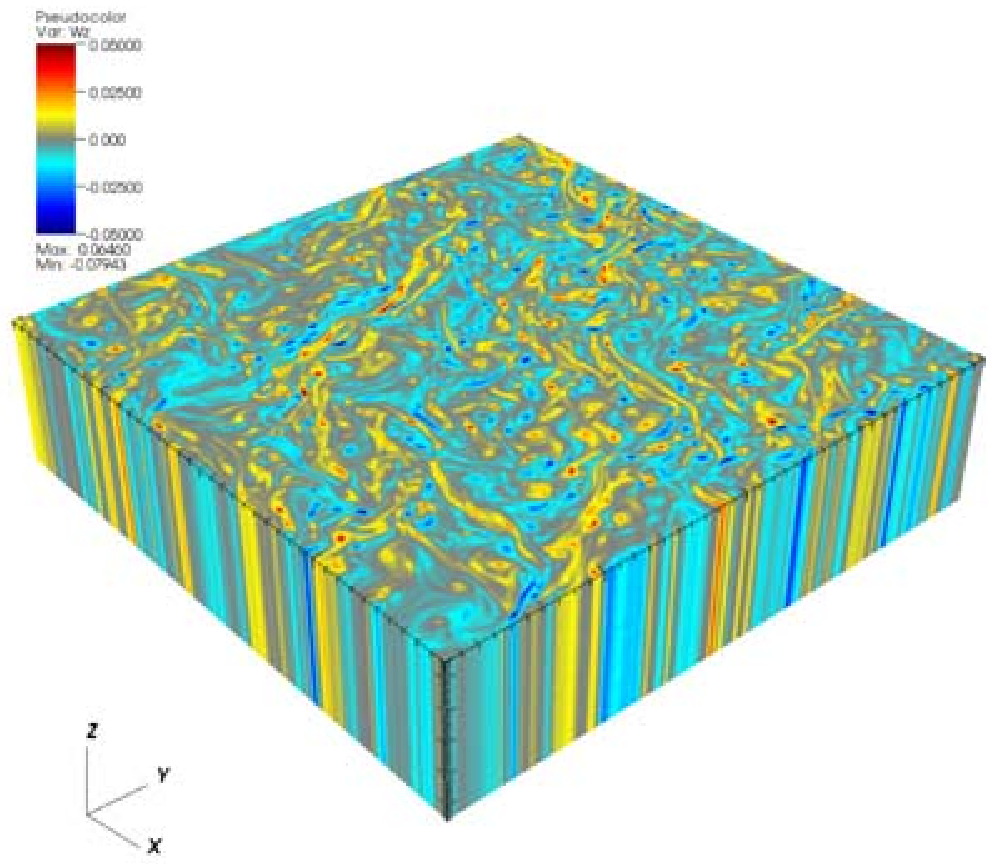}
    \includegraphics[width=0.45\textwidth]{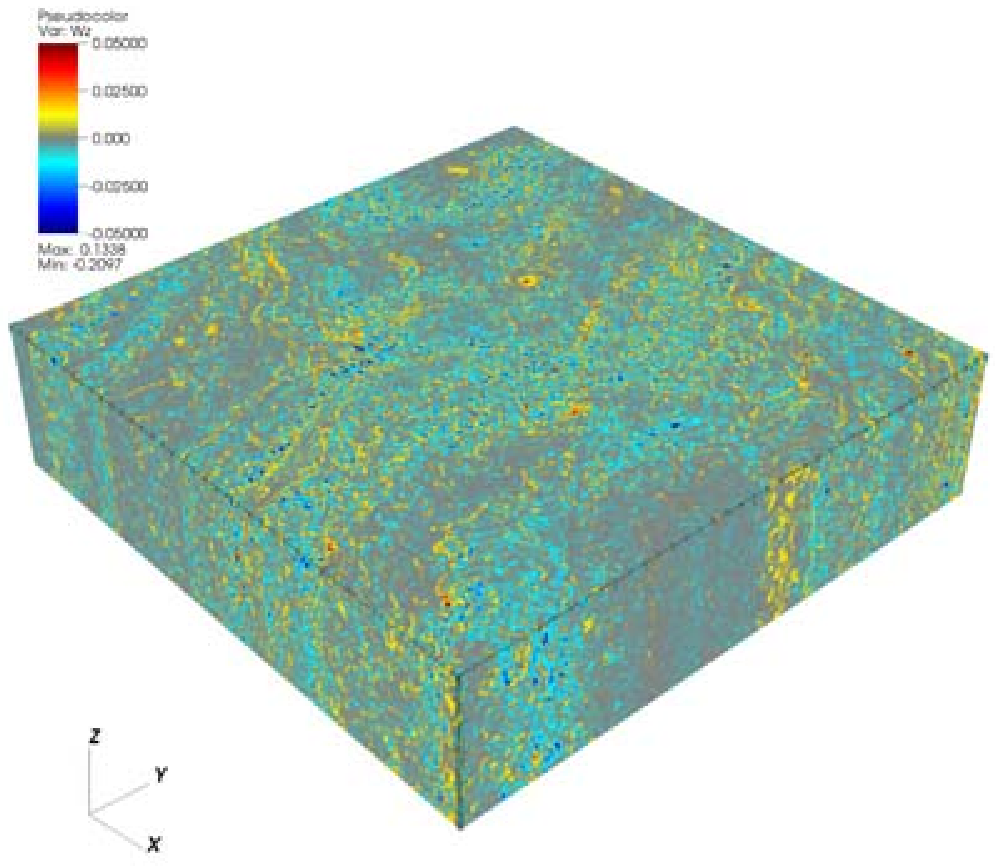}
    \\[-10pt]
    \caption{Slices of the $z$ component of vorticity $W_z$ for the $128^2\times 32$, $256^2\times64$,
    $512^2\times 128$ and $1024^2\times 256$ 3D runs taken at $\Omega t=5$.}
    \label{diffresol}
\end{figure*}

\subsection{Random Vorticity in Three-Dimensions}

Using numerical resolutions of $N=128$, $256$, $512$ and $1024$,
we have computed the evolution of random vortical perturbations in
3D.  We use the identical initial velocity field, generated by the
identical Fourier spectrum, as in the 2D case and simply project
it uniformly over the third dimension.  Thus, $W_z$ is independent
of $z$ in our initial conditions, and therefore we study the
evolution of vertical vortex columns in 3D.  To break symmetry in
the third dimension, we add small amplitude, random zone-to-zone
vertical velocity perturbations with maximum amplitude $0.05c_s$.

Figure \ref{fig3DRan} shows snapshots in the evolution of $\delta
W_z$ taken from the $N=512$ simulation at times of $\Omega t = 0,
10, 20$ and 60. Note the pattern of $W_z$ in the x-y plane at
$t=0$ is similar to the first panel of figure \ref{fig2DRan}.
Comparing the evolution to the 2D case, it is clear the vorticity
decays much more rapidly. Vertical symmetry is maintained until
$\Omega t =20$, at which point large fluctuations are present as a
function of vertical position $z$.  By $\Omega t=60$ the initial
vortex columns have disintegrated into a complex and intertwined
network of filaments that show no symmetries.  The small scale
structure introduced by the break up of vortex columns likely is
the cause of the rapid decay.  It is well known that columnar
vortices are subject to elliptical instabilities (e.g., Kerswell
2002), thus the destruction of the vertical vortex tubes observed
in figure \ref{fig3DRan} is not surprising.
%In particular, columnar vortices whose vertical extent exceeds their
%horizontal extent are easier to be destroyed by elliptical
%instabilities.
The breakdown of vortex columns has been reported in numerical
simulations (e.g., BM), where the initial configurations of
vortices are of known analytical forms and the growth rate can be
measured for isolated vortices. Due to the complicated vortex
dynamics in our simulation, i.e., background shear, vortices
interacting and merging, it is difficult to quantify the
destruction of vortices in terms of elliptical instability. Yet
the underlying physics should be similar. The elliptical
instability should exist for all two-dimensional elliptical
streamlines, where 3-dimensional Kelvin-mode disturbances become
unstable when resonating with the underlying strain field (e.g.,
Kerswell 2002). The result of the elliptical instability is to
break down the vortex column into small-scale structures, which is
what we find in our simulations.

Figure \ref{3Drandom} plots the same boxcar smoothed time
evolution of $\alpha$ and $E_{{\rm K}}$ from the 3D simulations,
along with the $N=2048$ 2D results (heavy solid curves) for
reference.  The plots demonstrate how much more rapid the decay of
stress and energy is in 3D compared to 2D. The evolution of stress
and $E_{\rm K}$ is a rapid exponential decay (from $\Omega t\sim
10-20$) followed by a slower power-law decay. The exponential
decay phase is probably associated with the breakup of the
vortices and the power-law decay phase is probably associated with
the decay of 3D hydro turbulence. The residual volume averaged
shear stress is at least one order of magnitude smaller than that
of the 2D case.

Perhaps the most striking aspect of the late time evolution of
$\alpha$ and $E_{{\rm K}}$ in figure \ref{3Drandom} are the small
amplitude (linear) oscillations.  By Fourier analysing the
complete (rather than boxcar averaged) history of $\alpha$ and
$E_{{\rm K}}$, we find a strong peak at a characteristic frequency
of $0.3\Omega$.  Similar, but less strong, frequency peaks are
also seen at late times in the 2D runs, although at a frequency of
$1.5\Omega$.  This frequency depends on boxsize, a 2D box with
dimensions $L_x=2H$ and $L_y=4H$ has a characteristic frequency of
the late time oscillations of $0.75\Omega$. Since the oscillations
are such small amplitudes, we conclude they must be related to
linear modes in the shearing sheet (Balbus 2003).  In 2D, only
spiral density waves are possible, in 3D both acoustic and nearly
incompressible inertial waves are present. The characteristic
frequency expected depends on the azimuthal wavenumber $m$, as
well as the ratio of the vertical to radial wavenumber for
inertial waves (e.g., Balbus 2003). Since they are nearly
incompressible, well-resolved (more than 16 grid points per
wavelength) inertial waves should decay very slowly in our
simulations. The observed frequencies are consistent with low-$m$
linear waves as the origin of the oscillations\footnote{The
characteristic frequency of inertial waves is given by $\omega -
m\Omega = \sqrt{k_z^2/(k_z^2+k_R^2)}\Omega$ (e.g., Balbus 2003).
So the observed low angular frequency oscillation ($\omega\sim
2\pi\times0.3\Omega $) could originate from inertial waves with
angular frequency $\omega = (m+\sqrt{k_z^2/(k_z^2+k_R^2)})\Omega
\sim 1.88\Omega$, which implies $m$ could be 1.}.

Figure \ref{diffresol} shows snapshots of $W_z$ at the same time
$\Omega t =5$ at four different resolutions; $N=128, 256, 512$ and
$1024$. No convergence in the spatial structure of $W_z$ is seen
with resolution. Instead, entirely different structures are
evident the same time at different resolutions.  In the lowest
resolution case $N=128$, the pattern of $W_z$ is very smooth, with
the remaining fluctuations fairly symmetric in $z$.  The amplitude
of the fluctuations increases with resolution, with the vertical
structure remaining symmetric up to the $N=512$ case.  However, at
$N=1024$, vertical symmetry is broken, and the vorticity has
disintegrated into small scale filaments similar to those observed
in the last panel of figure \ref{fig3DRan} for the $N=512$ case.
The lack of convergence in $W_z$ is quite striking. The higher
resolution simulations are able to capture higher wavenumber modes
of the elliptical instabilities that lead to the destruction of
vertical columns. The presence of these modes leads to destruction
of the columns at an earlier time at higher resolution. This trend
accounts for most of the difference in the structure between
different resolutions in figure \ref{diffresol}. The overall trend
that dynamical instabilities of the vortex columns destroys the
vertical symmetry and results in more rapid decay of $\alpha$ and
$E_{\rm K}$ is ubiquitous at every resolution.

\section{SUMMARY AND CONCLUSIONS}

Using the 3D version of the Athena code (Gardiner \& Stone 2005; 2006),
we have carried out hydrodynamical simulations of the evolution of
both planar waves and random vortical perturbations in
unstratified Keplerian disks using the shearing sheet
approximation. Our results can be summarized by the following four
points.

(1) Our numerical methods reproduce the evolution of the amplitude
of both compressible and incompressible (vortical) plane waves
predicted by linear theory with negligible dissipation as long as
the numerical resolution is 16 grid points per wavelength or
larger.  At lower resolution, the wave amplitude is smoothly
damped.  Significantly, there is no evidence of aliasing of
trailing into leading waves as they are sheared down to the grid
resolution.  We have demonstrated the amplitude error in plane
waves in the shearing sheet converges at better than second-order
with our methods.

(2)  We have shown that incompressible plane waves become KH
unstable and are destroyed when $\mid W_{\rm
max}\mid\gtrsim\Omega$,
%$\mid k_x \delta v_y\mid \gtrsim\Omega$,
the condition that the growth rate of the KH instability exceeds
the angular velocity. Whether the wave becomes KH unstable only
depends on the initial vorticity amplitude in the initial
conditions. In this case the planar shearing wave amplitude is
damped well before it gets amplified by the transient growth
mechanism.

(3) In 2D, the evolution of large amplitude random vorticity perturbations
follows the results reported by others
(Godon \& Livio 1999, 2000; Umurhan \&
Regev 2004; JGb).  In particular, coherent, large-scale (horizontal
extent larger than the scale height), anticyclonic
vortices survive and decay slowly.  At late times the dimensionless
angular momentum flux is of the order $10^{-3}$ averaged over
$100<\Omega t<200$, consistent with JGb.  This flux is dominated
by the residual motions in the large-scale anticyclonic vortices.

(4) In 3D, the evolution of large amplitude random vorticity
perturbations is similar to the 2D case, except the elliptic
instabilities destroy vortex columns whose vertical extent exceeds
their horizontal extent. This greatly increases the rate of decay
of kinetic energy and stress. The resulting volume averaged shear
stress $\alpha$ at late-times is at least one order of magnitude
smaller than that for the 2D case, and is probably associated with
linear amplitude, low-$m$ inertial waves that remain in the box,
and which decay slowly.

The destruction of planar vortical waves by the KH
instability has implications for transient amplification as a
means to drive strong turbulence.
%{\it The evolution of $\mid
%k_x\delta v_y\mid$ from leading to trailing is given by equation
%(\ref{KH_evo}), which first decreases with time in the leading
%phase and attain its minimum value of zero at $t_{{\rm max}}$ when
%$k_x=0$; then increases with time in the trailing phase.  Thus, it
%is possible to initialize leading waves which are KH stable, and
%which remain so all the way through peak amplification. However,
%even in this case $\mid k_x\delta v_y\mid$ increases monotonically
%in the trailing phase, and the wave may eventually become
%unstable. It is also possible to initialize leading waves which
%are KH unstable, but because $\mid k_x\delta v_y\mid$ decreases in
%the leading phase, these waves may become stable and survive
%through peak amplification; eventually these waves will be
%destroyed in the trailing phase, however. (I will modify it
%later)} I have commented out this paragraph since it is irrelevant now (Yue)
Although this might appear attractive as a feedback mechanism to
drive turbulence (Chagelishvili et al. 2003; Yecko 2003; Afshordi,
Mukhopadhyay, \& Narayan 2005), we find instability destroys the
wave and results in rapid decay of kinetic energy, rather than
re-seeding new leading waves. We speculate that most of the energy
in the vortical wave is converted into compressible modes by the
KH instability, which are not amplified by shear. Moreover, Balbus
\& Hawley (2006) found that these planar vortical wave solutions
are actually exact to all orders and cannot serve as a route to
self-sustained turbulence.

%For nonlinear amplitudes $\delta v_y \sim c_s$, the stability
%condition equation (13) reduces to $\mid k_x H\mid > 1$, where $H$
%is the scale height in the disk.  Thus, only nonlinear waves with
%large $\lambda_{x,0}$ initially, $\lambda_{x,0} \geq H$, may be
%stable. Similarly, one can show that the only waves which are
%initially KH stable, and which have nonlinear amplitudes at peak
%amplification, must have $k_y< 2\pi/H$.
% I also removed this paragraph, because it's less relevant (Yue)

Our results show no evidence for sustained hydrodynamical
turbulence in the shearing sheet.  While there is non-zero
transport, it is associated with large scale vortices that are
introduced in the initial conditions, or which emerge from mode interactions
during the nonlinear decay phase.  At late times the Reynolds stress
is oscillatory, which may indicate long-lived linear
amplitude inertial waves also contribute to the time-averaged shear.

To understand the relevance of large-scale vortices to the dynamics
of astrophysical disks, it will be important to understand how they
are generated and destroyed. BM have shown that such vortices can be
generated in stratified disks, but given the importance of compressible
waves on the decay of vortices (JGb), it is important to investigate
vortex production and evolution in fully compressible stratified disks.

\section*{Acknowledgments}
We thank Steve Balbus, Charles Gammie, Bryan Johnson, Jeremy
Goodman, John Hawley, and Ramesh Narayan for stimulating
discussions.  We thank the referee, Thierry Foglizzo, for a report
that led to significant improvements in the manuscript.
Simulations were performed on the IBM Blue Gene system at
Princeton University, and on computational facilities supported by
NSF grant AST-0216105.

%\begin{figure} \centering
%\includegraphics[scale=0.6]{N256_2D_3D_raw.eps}
%\caption{} \label{fig2D3Draw}
%\end{figure}

\end{document}